\documentclass[sigconf]{acmart}

\AtBeginDocument{%
  }

\copyrightyear{2022}
\acmYear{2022}
\setcopyright{rightsretained}
\acmConference[UIST '22]{The 35th Annual ACM Symposium on User Interface Software and Technology}{October 29-November 2, 2022}{Bend, OR, USA}
\acmBooktitle{The 35th Annual ACM Symposium on User Interface Software and Technology (UIST '22), October 29-November 2, 2022, Bend, OR, USA}
\acmDOI{10.1145/3526113.3545681}
\acmISBN{978-1-4503-9320-1/22/10}

\acmSubmissionID{6474}

\begin{document}

\newcommand{\rev}[1]{\textcolor{black}{#1}}
\title[Scholastic: Graphical Human-AI Collaboration for Inductive and Interpretive Text Analysis]{Scholastic: Graphical Human-AI Collaboration\\for Inductive and Interpretive Text Analysis}

\author{Matt-Heun Hong}
\orcid{0000-0003-3169-9654}
\affiliation{%
  \department{ATLAS Institute}
  \institution{University of Colorado Boulder}
  \city{Boulder}
  \state{CO}
  \country{USA}
}

\author{Lauren A. Marsh}
\affiliation{%
  \department{Department of Applied Mathematics}
  \institution{University of Colorado Boulder}
  \city{Boulder}
  \state{CO}
  \country{USA}
}

\author{Jessica L. Feuston}
\affiliation{%
   \department{Department of Information Science} \institution{University of Colorado Boulder}
  \city{Boulder}
  \state{CO}
  \country{USA}
}

\author{Janet Ruppert}
\affiliation{%
   \department{Department of Information Science} \institution{University of Colorado Boulder}
  \city{Boulder}
  \state{CO}
  \country{USA}
}

\author{Jed R. Brubaker}
\affiliation{%
  \department{Department of Information Science}  \institution{University of Colorado Boulder}
  \city{Boulder}
  \state{CO}
  \country{USA}
}

\author{Danielle Albers Szafir}
\affiliation{%
  \department{Department of Computer Science}  \institution{University of North Carolina\\ at Chapel Hill}
  \city{Chapel Hill}
  \state{NC}
  \country{USA}
}

\renewcommand{\shortauthors}{M.-H. Hong, L. A. Marsh, J. L. Feuston, J. Ruppert, J. R. Brubaker, and D. A. Szafir}

\begin{abstract}
Interpretive scholars generate knowledge from text corpora by manually sampling documents, applying codes, and refining and collating codes into categories until meaningful themes emerge. Given a large corpus, machine learning could help scale this data sampling and analysis, but prior research shows that experts are generally concerned about algorithms potentially disrupting or driving interpretive scholarship. We take a human-centered design approach to addressing concerns around machine-assisted interpretive research to build \textit{Scholastic}, which incorporates a machine-in-the-loop clustering algorithm to scaffold interpretive text analysis. As a scholar applies codes to documents and refines them, the resulting coding schema serves as structured metadata which constrains hierarchical document and word clusters inferred from the corpus. Interactive visualizations of these clusters can help scholars strategically sample documents further toward insights. 
\rev{\textit{Scholastic} demonstrates how human-centered algorithm design and visualizations employing familiar metaphors can support inductive and interpretive research methodologies through interactive topic modeling and document clustering.}
\end{abstract}

\ccsdesc[300]{Computing methodologies~Natural language processing}
\ccsdesc[500]{Human-centered computing~Visual analytics}
\ccsdesc[300]{Human-centered computing~Collaborative and social computing systems and tools}

\keywords{qualitative research, interpretive research methods, interactive topic modeling, interactive document clustering, human-AI collaboration, visual analytics, text data}

\begin{teaserfigure}
  \includegraphics[width=\textwidth]{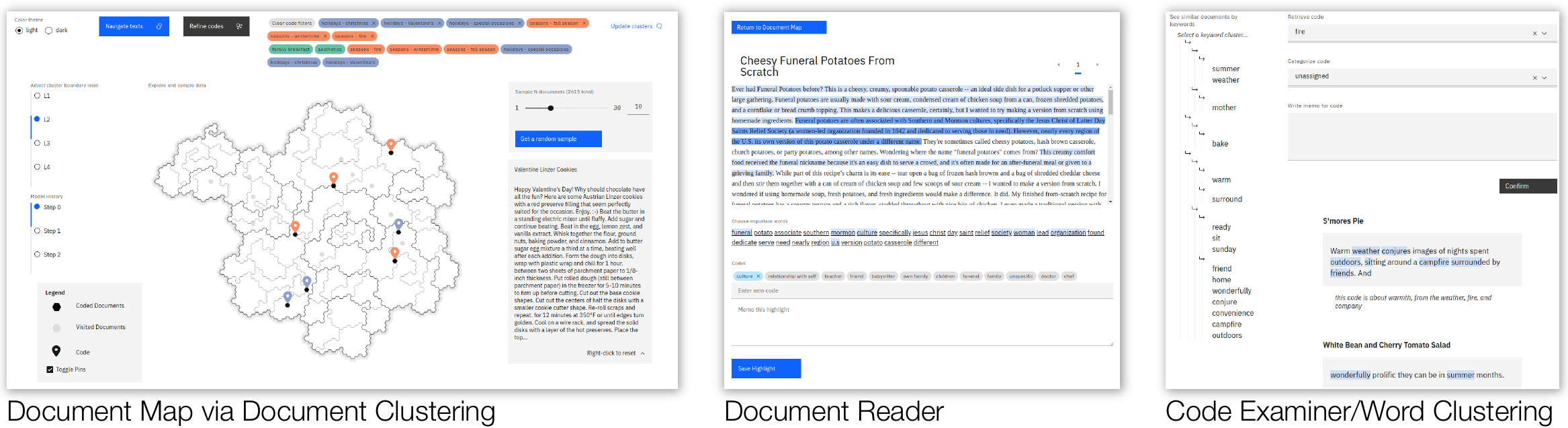}
  \caption{The \textit{Scholastic} interface. \textmd{\textit{Scholastic} aims to support core elements of an interpretive qualitative analysis workflow for analyzing text documents using visual analytics and interactive machine learning. The system is comprised of three views that afford a variety of strategies for document sampling, applying codes to passages within individual documents, and refining and categorizing codes.}}
  \Description{interface navigation flow}
  \label{fig:teaser}
\end{teaserfigure}
\maketitle

\section{Introduction}
\label{intro}
Modern social science depends heavily on analyzing text data, such as interviews, written logs, or social media archives. Researchers may infer patterns from raw texts using statistical topic modeling \cite{Blei2003}. Alternatively, researchers may employ interpretive methods, clustering texts based on an in-depth reading of the data to apply codes and group codes into categories, which are then iteratively refined until meaningful themes emerge \cite{Braun2006}. The two approaches represent trade-offs between efficiency and efficacy. The labor-intensive interpretive methods may not scale to a large corpus (e.g., collections of online blog posts or tweets), requiring researchers to only examine (an often random) sample of texts \cite{Gauthier2022}. While statistical models can rapidly process a large corpus, their reliance on purely statistical patterns in the absence of expert knowledge can sacrifice semantics for scale, leading to findings that may insufficiently address critical research questions \cite{Chen2018a}. 

\rev{Visualizing topic models and document clusters could help interpretive scholars explore both the breadth and depth of content in a corpus \cite{Baumer2020a}. However, our prior interview studies \cite{Jiang2021, Feuston2021} discovered that interpretive scholars were largely skeptical about using machine learning to support their analyses, raising concerns about algorithms driving 
or replacing human expertise and biasing the analysis process (\S\ref{objectives}). 
Recent human-AI collaboration tools 
can capture expert knowledge as
input to refine statistical text models \cite{El-Assady2020}. However, using these tools 
would require interpretive scholars to work outside of typical workflows where they freely and iteratively apply codes to documents. 
In this work, we introduce \textit{Scholastic}, a visual analytics tool that
instead explores a machine-in-the-loop \cite{green2019disparate,green2019principles} approach to interpretive research, where scholars analyze text to generate codes and categories that also serve as goal-oriented user input to models that scaffold, rather than replace, human sensemaking. }
 \textit{Scholastic} aims to build on the strengths of interactive topic modeling and document clustering for helping organize text data at scale while minimizing disruptions to a focused qualitative analysis workflow.

\rev{When scholars sample documents to examine and apply codes to a passage within a document, the code label becomes both a meaningful unit of information as well as an organizational tool for re-examining relevant passages from the corpus \cite{miles2013qualitative} to refine the scholar's coding schema. We consider two additional conceptualizations of codes as: 1) meaningful human input for the text model to learn from (\S\ref{algorithm}.1), and 2) interactive filters to visualize the distribution of emerging knowledge across clusters (\S\ref{interface}).
Hierarchical document and word clusters generated by an interactive topic modeling algorithm are depicted using interactive geographical treemaps \cite{Auber2011} and indented trees \cite{Kim2020}, drawing on familiar visual metaphors while supporting 
evolving
strategies for information foraging. Scholars can apply and iterate on individual codes using the raw text, moving freely between clusters and text as their analysis develops.} 

\rev{\textit{Scholastic} is the result of a multi-phase co-design process with interpretive scholars, machine learning researchers, and visualization scientists. 
Our prototype 
provides preliminary insight into the vision of 
incorporating interactive ML within the data sampling and sensemaking loops of a qualitative analysis workflow given a large corpus (e.g., online blog posts).
Our primary contribution, \textit{Scholastic}, is a visual analytics tool that supports interpretive data analysis at scale, which comprises:}
\rev{
\begin{itemize}
    \item An interactive word and document clustering algorithm that 
    incorporates evolving codes and categories as 
    model constraints, 
    \item Reading, coding, and categorization tools familiar to interpretive scholars,
    \item Interactive cluster visualizations that support both breadth-first exploration of and depth-first search for relevant documents, and 
    \item A characterization of the design needs for graphical tools supporting qualitative analysis workflows.
\end{itemize}
We conducted a formative user study of the tool (\S\ref{evaluation}), focusing on its usability for sampling and coding processes.
}

\section{Background}
\label{background}
\subsection{Inductive and Interpretive Text Analysis}
Interpretive research methods for making sense of text data include thematic analysis \cite{Braun2006} and grounded theory analysis \cite{Corbin1990}. One shared thread between these approaches is a principled method for data collection (or sampling items when given an existing corpus) given a research question. Given the collected (or sampled) dataset and in the absence of prior relevant knowledge about a population under study (`data-driven’ as opposed to `theory-driven’ analysis \cite{Braun2006}), the process of applying codes to texts and iteratively categorizing those codes \cite{miles2013qualitative} is the central process of inductively modeling meaningful patterns (`surfacing themes’) within interpretive analysis.

The deluge of data available has in some ways made data collection easier, but data items must still be sampled from a corpus, and applying codes can be laborious for even small sets of interviews or ethnographic data \cite{Muller2016}. Popular tools like MaxQDA\footnote{maxqda.com} or NVivo\footnote{qsrinternational.com} provide environments in which analysts can manage texts and codes, but these tools can only produce basic summary statistics like code counts. The visualization features in our system prototype are designed to support the data sampling process for interpretive analysis, but also provide coding and categorization functionalities to 1) provide scholars an effective algorithmic support tool to think with and 2) collect user input for an interactive ML algorithm.

\subsection{Using Topic Models for Qualitative Research}
Epistemological discussions surrounding how machine learning could be leveraged for interpretive research \cite{Baumer2017, Chuang2015} have noted similarities between coding in qualitative analysis and topic modeling: both approaches share the goal of iteratively inferring models without prior labels. 
Topic models
such as Latent Dirichlet Allocation (LDA) \cite{Blei2003}
output probabilistic clusters of words based on their co-occurrence patterns within documents. 
One application of topic models is to facilitate document clustering \cite{Xie2013}. Boyd-Graber et al. \cite{Boyd-Graber2017} review the use of topic models 
across digital humanities and social sciences.
These works focus on refining topic models (e.g., by introducing new random variables \cite{Paul2010}) and statistically validating them (e.g., using posterior predictive checks \cite{Mimno2011}) to output best-fitting statistical models summarizing text corpora.

On the other hand, the high probability words in each topic may also be read as `themes' that provide an overview of a text corpus \cite{Baumer2017}. 
When interpreted this way, topic models can sometimes suggest overlooked codes and categories or open up paths to
other meaningful documents.
For example, through human
interpretation and sampling via topic models, Nelson \cite{Nelson2020} analyzed suffrage and feminism movements in New York City and Chicago
to characterize how these two local movements differed in their guiding political principles.
Although refining and validating topic models may appeal to qualitative scholars for their statistical power, our research team and the qualitative scholars we interviewed \cite{Jiang2018a, Feuston2021} agree with the views of Nelson and Grimmer \& Stewart \cite{Grimmer2013} that topic models can be a tool for assisting---but not superceding---human induction when conducting data-driven interpretive analysis. Still, \textit{how} models should be integrated into
human-AI
collaborative workflows remains an open question within interpretive scholarship.

\subsection{Interfacing with Text Models}
Text models can be interpreted and refined using graphical tools. Termite \cite{Chuang2012} represents the probability distributions of topic models with matrix-based representations. 
Topicalizer \cite{Baumer2020a}, TOME \cite{Klein2015}, and Serendip \cite{Alexander} follow a more human-centered design approach for presenting topic models for qualitative research or digital humanities.
UTOPIAN \cite{Choo2013} and ArchiText \cite{Kim2020} allow users to improve topic models using various interactions, including merging and splitting topics and removing words from topics. Lee et al. \cite{Lee2017} surveyed and evaluated these strategies, recommending eight useful interactions for topic refinement. In contrast, our algorithm design is closest to Yang et al. \cite{Yang2015a} who incorporated expert knowledge as `must-link' or `cannot-link' constraints for LDA via factor graphs. 

Visual analytics approaches also enable users to understand how their interactions change clustering outputs. For example, iVisClustering \cite{Lee2012a} allows users to see how adjusting topic models impacts document clustering outputs. Endert et al. \cite{Endert2012} introduces \textit{semantic interaction} for directly updating document embeddings. Semantic Concept Spaces \cite{El-Assady2020} helps analysts incorporate expert knowledge about data semantics into topic models through direct manipulation. 
Our algorithm design takes the goal of incorporating data semantics further by gathering user input given clear research objectives and methodologies while surfacing codes and categories. 
Related tools for coding documents include Overview \cite{Brehmer2014}, an investigative journalism tool for sampling and categorizing documents given hierarchical clusters. 
Aeonium \cite{Drouhard2017a} is a collaborative coding tool that helps identify disagreements between scholars via an SVM classifier.
Chandrasegaran, et al. \cite{Chandrasegaran2017c} leverages
NLP to highlight keywords 
across documents to support sensemaking within interpretive scholarship.  
Our tool provides a contrasting view on supporting interpretive scholarship with visual analytics by focusing on document sampling and interactive modeling in inductive qualitative workflows.

\begin{figure}[h]
  \centering
  \includegraphics[width=\linewidth]{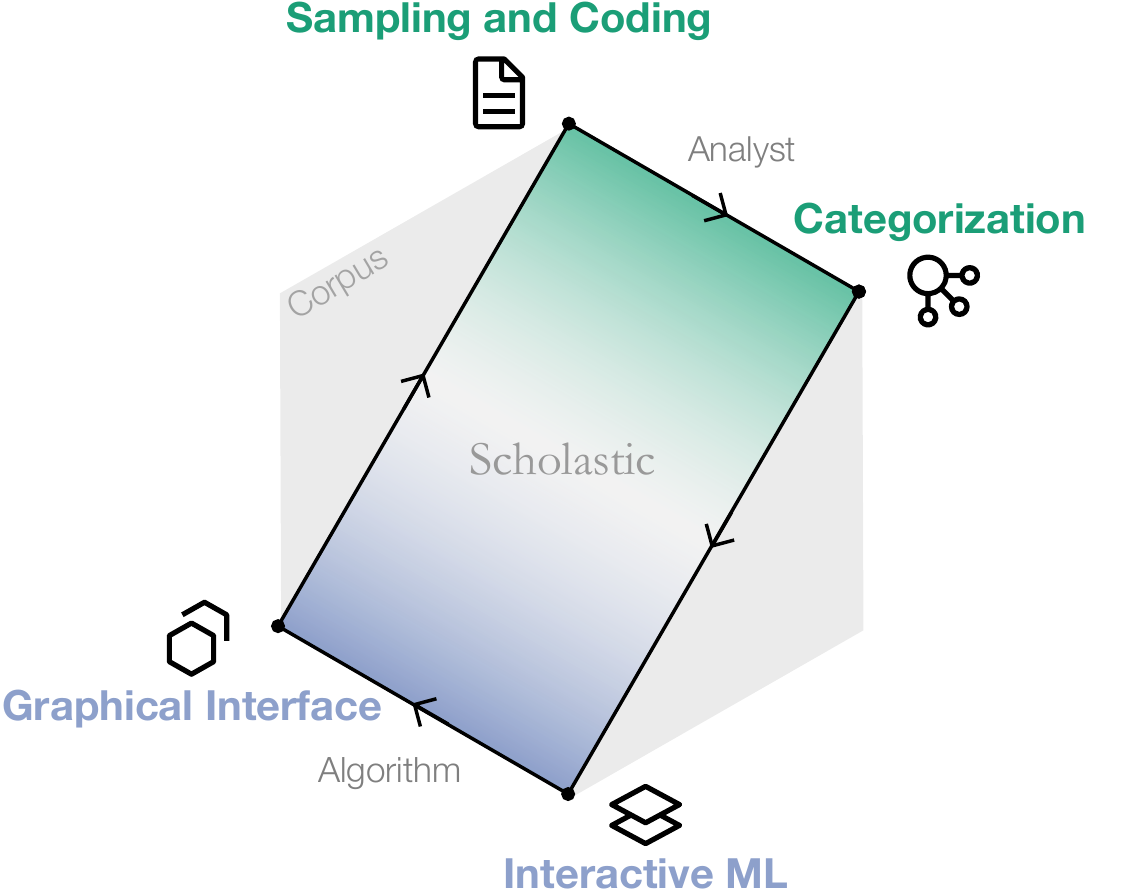}
  \caption{\textit{Scholastic}'s machine-in-the-loop workflow. \textmd{
  Sampling, coding, and categorization are performed by the analyst. The algorithm  can then hierarchically cluster the documents and words using the 
  codes and categories as constraints, and represent the outputs with interactive visualizations for the analyst to sample more documents. The analyst retains agency over when the algorithm will perform these tasks. 
  }}
  \Description{machine-in-the-loop workflow}
  \label{fig:workflow}
\end{figure}

\section{Design Objectives}
\label{objectives}
We characterized four key design considerations for interpretive analysis
based on recent interview studies conducted by the research team \cite{Jiang2021, Feuston2021} as well as internal discussions with experts throughout the design process.
Popular methodologies such as thematic analysis and grounded theory analysis share several common components. Most notably, they define processes for selecting or collecting data, inferring codes or categories from that data, and organizing and refining those codes to build knowledge.

On the other hand, researchers across fields have different ways of framing the interpretive analysis process.
To resolve potential ambiguities in terminology, we refer to an individual document as a \textit{data item}, a collection of documents as a \textit{data sample}, and the entire collection of documents as a \textit{corpus}. 
\textit{Coding} is the act of applying labels to text passages,
and \textit{categorizing} is the act of collating codes. \textit{Memoing} is recording notes \cite{miles2013qualitative} to later recall reasons for coding passages or to log additional expert insights.


\subsection{Consideration 1: Supporting Serendipity}
\rev{Interviews from Jiang et al. \cite{Jiang2021}
stressed that machines should not 
lead data sensemaking in interpretive analysis. Scholars felt that traditional algorithms that output
potentially immutable data summaries could bias knowledge generation. 
For example, they felt that traditional algorithmic approaches could overly constrain how analysts see the data (e.g., by asserting an algorithmic definition of the most ``meaningful'' patterns in data or creating anchoring biases), bias interpretation by dictating the most ``important'' terms associated with document clusters prior to analysis (e.g., by labeling clusters according to the most frequent or highest probability terms), and put the algorithm in control of the data analysis workflow (e.g., by limiting how analysts can compare documents across clusters).
They also felt that
``maybe [the machine] could make suggestions, but even then I don't know if I want it because it doesn't know what my research questions are." 
From the interpretivist view, text data has inherent ambiguities related to semantics, and experts wanted to resolve those ambiguities themselves to support their own knowledge generation: 
``It's really satisfying ... it's those kinds of exciting Eureka moments that make research kind of worth it.”}

\rev{AI could instead
support serendipitous moments where the human analysts resolve ambiguities about the data to foster moments of insight. 
Therefore, our goal was to 
support curiosity about the data by scaffolding data sampling with a mixed-initiative visual analytics tool. 
Muller et al. \cite{Muller2016} also previously suggested that cluster models can provide human scholars with alternative representations of data with which to refine codes.
While past approaches have used NLP algorithms to identify 
important segments across texts 
to explicitly drive insight generation within qualitative analysis \cite{Chandrasegaran2017c}, 
researchers we spoke with felt this approach shifted too much analytical power to the algorithm to the detriment of the analysis process.}



\subsection{Consideration 2: Right Place, Right Time}
\rev{Feuston \& Brubaker \cite{Feuston2021} described various computational subsampling strategies used by scholars, including simple random sampling. 
While reticent to use automation in coding or categorization, scholars were willing to delegate data sampling to algorithms often as a matter of practicality: corpora are often too large to code all data items.
One scholar had used cluster overview visualizations from semantic network analysis alone to sample data items; another used classifiers grounded in keywords related to categories they had developed to identify similar data items.
While both approaches illustrate algorithmic sampling strategies, the latter approach combines human expertise with automation.}

\rev{Many scholars felt that 
AI was only appropriate after they had made some analytic progress, as in the second analyst's keyword-based approach.
Cluster models often rely on word-document co-occurrence matrices that privilege frequently occurring words \cite{Steyvers2010}. Frequency is not necessarily integral to qualitative methodologies. Data patterns of interest can be sparsely scattered, 
so if text models are used at all, they should incorporate human inputs in addition to co-occurrence information.
The updated clusters can then be used to guide further sampling (adhering to the \textit{constant comparative method} in grounded theory).
We note the potential caveat that human selection bias 
also poses a
challenge to generalizability within qualitative research \cite{collier_mahoney_1996}. }

\subsection{Consideration 3: Using Familiar Paradigms}
Qualitative researchers often only rely on the most basic features of software \cite{wiedemann2013opening}. Scholars interviewed by Jiang et al. \cite{Jiang2021}
attributed this to the overall difficulty of using complex qualitative analysis
tools like MaxQDA or NVivo: ``[qualitative analysis tools should not be] like the NVivo type, where I have to really learn a lot of it." Such complexity was perceived as getting in the way of their analysis. Many scholars used Google Docs or post-it notes to manage codes. Sensemaking about computational tools can inadvertently hinder or misguide sensemaking about data using those tools: ``Using any tools, I think it gets in the way of the analysis... I think the focus then inevitably becomes on the tool and how I can manipulate and push data in order to make it appropriate for the tool." 
Building on these observations, interpretive analysis tools should, whenever possible, leverage familiar visual and interaction paradigms to help analysts retain their focus on the data rather than on navigating the tool.

\subsection{Consideration 4: Overlaying Visualizations With Codes and Categories}
Scholars interviewed by Feuston \& Brubaker \cite{Feuston2021} described how ``it might be interesting to compare and contrast" the analyst-inferred codes with machine-inferred clusters 
to help refine codes and categories. This desire was echoed by scholars in Jiang et al. \cite{Jiang2021} who frequently requested visualization features that allowed comparisons across clusters and codes: ``I want to be able to say, okay, all the people I've talked to who identify as queer, how did they feel about capitalism? I want to be able to do a cross-sectional analysis on multiple codes and domains.” This comparison may be accomplished by incorporating visual overlays of the applied codes and categories onto cluster visualizations, situating both human-induced and machine-inferred models in the same space. 

These comparisons also could foster collaborative interaction between human and automated analyses. For example, once a machine has learned from human input, the visualizations could guide the user toward sets of related documents or codes. 
An expert in Jiang et al. \cite{Jiang2018a} indicated that it would be useful ``if there was some sort of learning algorithm, for example, that would suggest... other quotes that were similar to that one."

\subsection{Summary of Design Objectives}
The above considerations indicate a need for 
human-AI collaboration 
in qualitative analysis, such that AI is embedded \textit{within} an interpretive research workflow and adapts to evolving codes and categories.
System features should support: 

\begin{enumerate}
 \item \textit{Insight Generation and Retention:} We aim to develop an interface for memoing, developing codes and categories, and revisiting coded documents to help people generate knowledge by creating, applying, and refining their coding schema with intelligent and transparent system support. 
 \item \textit{Random Sampling:} We aim to enable a random subselection of documents in the absence of human codes to gather initial insights about the dataset.
 \item \textit{Strategic Cluster-Based Sampling:} We aim to incorporate cluster model visualizations that enable both breath-first exploration of the corpus or depth-first search for potentially meaningful data items. Overlaying codes and categories onto the cluster visualizations will allow scholars to compare the model outputs with their own coding schema. 
 \item \textit{Familiar Metaphors for Interactive Visualizations}: We aim to leverage familiar visual metaphors for representing text models such that sensemaking about the tool does not 
 inhibit sensemaking about the data.
 \item \textit{Interactive Models that Incorporate Expert Input:} We introduce an interactive clustering algorithm that can learn from parsimonious human input without disrupting their analysis by using codes and categories as additional data for the model.
\end{enumerate}

\section{Implementation}
\label{implementation}
\textit{Scholastic} is a web-based application built using Python and JavaScript. Interactive machine learning is implemented through the Python packages spacy, graph-tool, NumPy, and SciPy. The backend interface utilizes Flask, pandas, and the Google Sheets API. As requested by experts, user inputs (codes, categories, passages, keywords, and memos) are saved to Google Sheets to support the integration of their analysis into existing external processes. The frontend interface is implemented through the Svelte, Carbon Design, TopoJSON, and D3.js packages.

\begin{figure}[h]
  \centering
  \includegraphics[width=\linewidth]{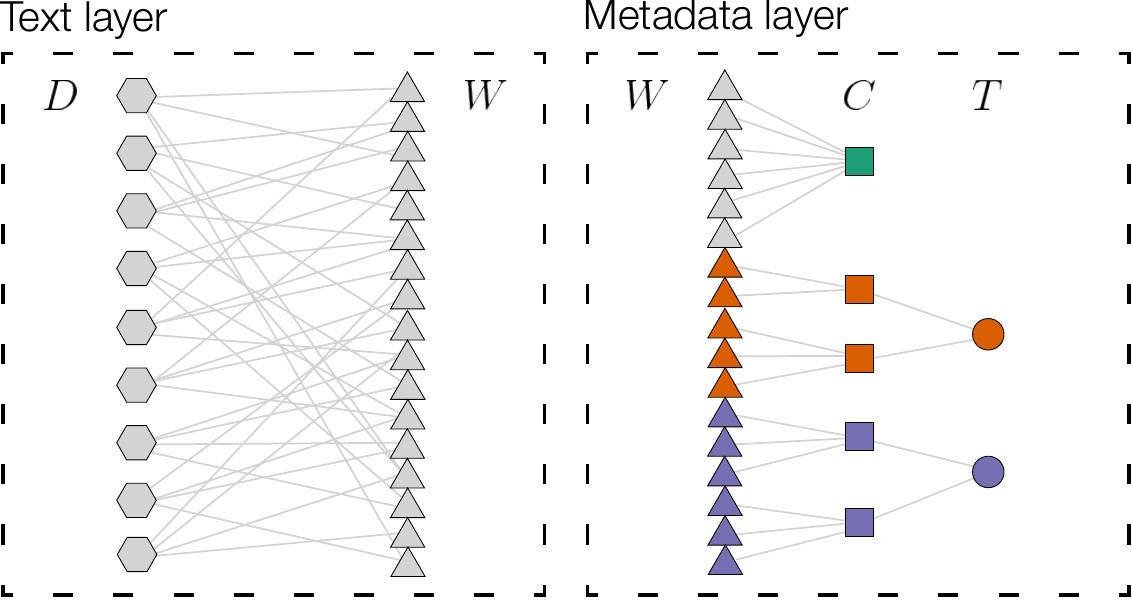}
  \caption{Undirected multilayer network representation of the corpus and human-induced metadata. \textmd{Note that this network allows parallel edges from the word nodes. The \textit{Text} layer is a bipartite network between the documents (hexagons) and words (triangles). 
  The \textit{Metadata} layer is a disconnected, tripartite network of words, codes (squares), and category tags (circles). The edges between words and codes will be input via passage highlighting and subsequent `in vivo' keyword selection, as described in \S\ref{sec:reader}. By default, words are assigned the \textit{non-keyword} code (green).}}
  \Description{network representation of corpus and user metadata input}
  \label{fig:layers}
\end{figure}

\section{Algorithm Design}
\label{algorithm}
\begin{figure*}[h]
  \centering
  \includegraphics[width=\textwidth]{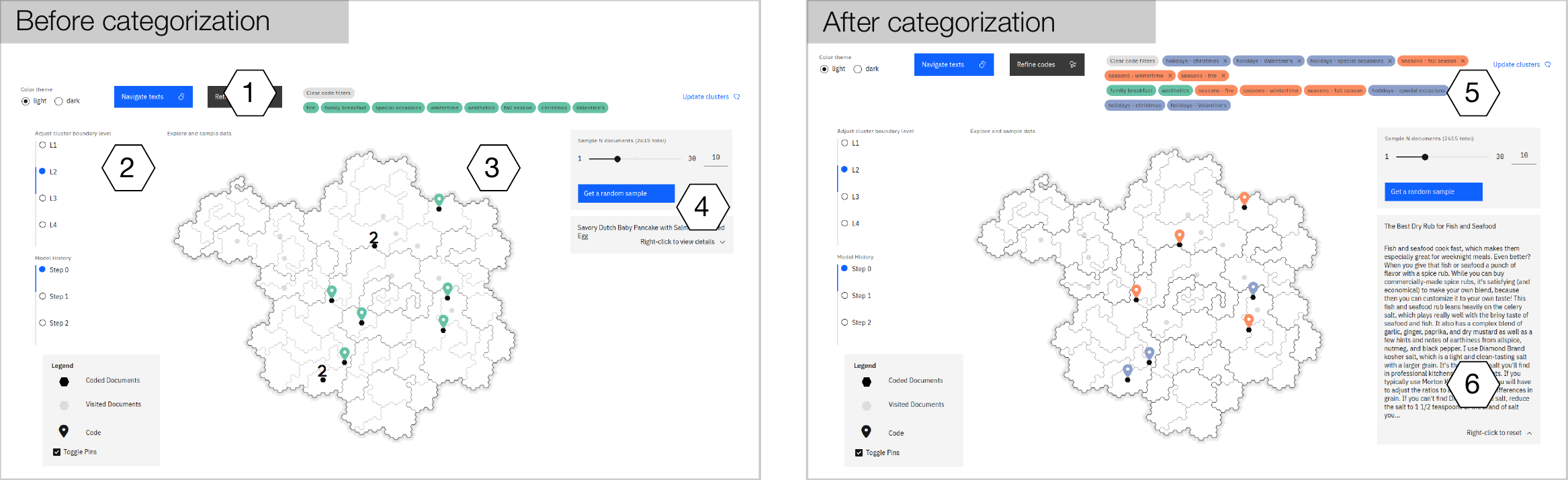}
  \caption{\textit{Scholastic}'s Document Map features and states. \textmd{The geographical treemap represents a hierarchical clustering of the corpus with color-coded pins summarizing applied codes (colors correspond to code categories). 
  A navigation bar (1) persists across all views of the interface, and
  the left sidebar (2) contains two sliders that can step through either the lowest level of the hierarchical clusters displayed or the history of the evolving model. The item legend component at the bottom of the left sidebar allows a scholar to remove all overlaid pins (3) and avoid visual occlusion. 
  Code filters within the navigation bar filter overlaid pins, which appear above documents if codes have been applied to passages within those documents.
  The random sampling component (4) generates a set of $N$ documents for the analyst to apply codes in the Document Reader. 
  Categorizing codes through the Code Examiner changes the colors of code tags and pin overlays (5).
  \rev{Code colors use D3.js's Set2 color scheme, applying one color per category. If there are more than eight categories, the color hues will duplicate as in commercial tools like Tableau, but the code filters still explicitly encode their categorization through their text labels.
  Hovering over each hexagon dynamically displays the corresponding document's title on the right sidebar; right-clicking the hexagon expands this component to show a content preview (6). Left-clicking a hexagon navigates the researcher to the Document Reader.}
  }}
  \Description{the document map is generated using hexagons to tile a surface, and used to sample documents by clicking on each hexagon.}
  \label{fig:document-map}
\end{figure*}

We adapt the hSBM approach to probabilistic word and document clustering by Gerlach et al. \cite{Gerlach2018} which uses a stochastic block model to detect communities in complex bipartite networks formed by text data.
A stochastic block model \cite{Holland1983} generates a random graph whose adjacency matrix representation is $A_{\mathrm{ij}}$ with probability $P(\mathbf{A} \mid \mathbf{b})$, 
where elements in the vector $b_{k}$
represent block membership assignments. 
In the context of text, each $A_{\mathrm{ij}}$ represents the number of times a word $w_i$ occurs in a document $d_j$, and $\mathbf{b}^W$ and $\mathbf{b}^D$ represent individual word and document blocks respectively. 
Given the marginal likelihood function 
defined in Gerlach et al. \cite{Gerlach2018}, the posterior distribution $P(\mathbf{b} \mid \mathbf{A})$ can be efficiently approximated using Markov Chain Monte Carlo (MCMC), which is then equilibriated to avoid local optima \cite{Peixoto2019}.

The informative priors on hSBM produce more heterogeneous mixtures than LDA while being completely non-parametric \cite{Gerlach2018}. 
Notably, the mixed-membership (`overlapping') version of hSBM, which outputs `soft' clusters of words (`topics'), 
significantly outperforms LDA topic models even on synthetic Dirichlet mixtures.
Additionally, even in the absence of stop-word removal, hSBM automatically detects clusters of stop-words which frequently occur across the corpus; it will also infer
the number of word and document clusters directly by sampling from
the
posterior distribution rather than requiring either the developer or researcher to specify a target number of clusters \textit{a priori}. 

In this work, we utilize the non-overlapping variant of hSBM.
Non-overlapping blocks partition the text data deterministically
(e.g., $P(\mathbf{b}^W_l|w_i)=1$ if word node $w_i$ belongs in a word cluster $\mathbf{b}^W$ in level $l$ and $P(\mathbf{b}^W_l|w_i)=0$ otherwise). In contrast to LDA, this forgoes the need for the system or its users to set a minimum probability threshold to obtain `hard' clusters from topic models.~\footnote{This partly motivates our avoidance in this paper toward referring to probabilistic word clusters as `topics'; interpretive scholars may also find the concept of `topics' difficult to disassociate from `themes.' } 
Given this bipartite model structure, hSBM infers hierarchical word and document block assignments $P(\mathbf{b}^W_l|w_i)$ and $P(\mathbf{b}^D_l|d_j)$ simultaneously.  

\subsection{Incorporating Codes and Categories \\as Metadata}
\label{sec:multilayermodel}
Adapting the multilayer hSBM introduced by Hyland et al. \cite{Hyland2021}, \textit{Scholastic} pairs word-document co-occurrence matrices with analyst-induced coding schema, adjusting clusters to reflect ongoing expert analyses.
To formulate these coding schema as constraints to cluster outputs, we re-frame the data types which characterize interpretive text analysis as follows:
\begin{description}
    \item[Documents] A corpus $D$  of document nodes $d_j$.
   \item[Words] A vocabulary $W$ of word nodes $w_i$.
   \item[Codes] Each $w_i$ in $W$ is classified by a code in $C$.
   \item[Categories] Each code in $C$ is classified by a category tag in $T$.
\end{description}

We represent these variables as an undirected multilayer network \cite{Kivela2014} with parallel edges (Figure 3), whose clusters can then be inferred using stochastic block models. Our network includes two layers: a \textit{Text} layer to capture co-occurrence patterns in text and a \textit{Metadata} layer for codes and categories generated by the human scholar. The \textit{Text} layer is a bipartite network with parallel edges between $W$ and $D$.
The \textit{Metadata} layer is a disconnected, tripartite network of words, codes, and category tags where parallel edges 
hierarchically partition words. 

When applied to multilayer networks, hSBM will simultaneously infer clusters across all layers. 
Since the same $W$ occurs in both the \textit{Text} and \textit{Metadata} layers, how words are clustered together will be identical across both layers \cite{Hyland2021}.
Thus, the partitioned nature of the
\textit{Metadata} layer 
enforces a constraint that keywords applied the same code 
must always be clustered together. 
However, the relationships of these keywords to other words in the \textit{Text} layer 
allows 
sets of keywords to be clustered with other non-keywords or 
with other keyword clusters. 

The edges between words and codes is inferred from passage highlighting and subsequent `in vivo' keyword selection used to apply codes to raw texts, as described in \S\ref{sec:reader}.
Note that every word in the corpus is always adjacent to a unique code.
If a word has 
not been coded, it remains by default adjacent to the \texttt{non-keyword} code.
After a scholar produces codes and categories, they can update the model on demand using a button in the interface (see \S\ref{interface}). The \textit{Metadata} layer is then remodeled and clusters reinferred with the new constraints that reflect the current analysis state. 

\section{Interface Design}
\label{interface}

\begin{figure*}[h]
  \centering
  \includegraphics[width=\textwidth]{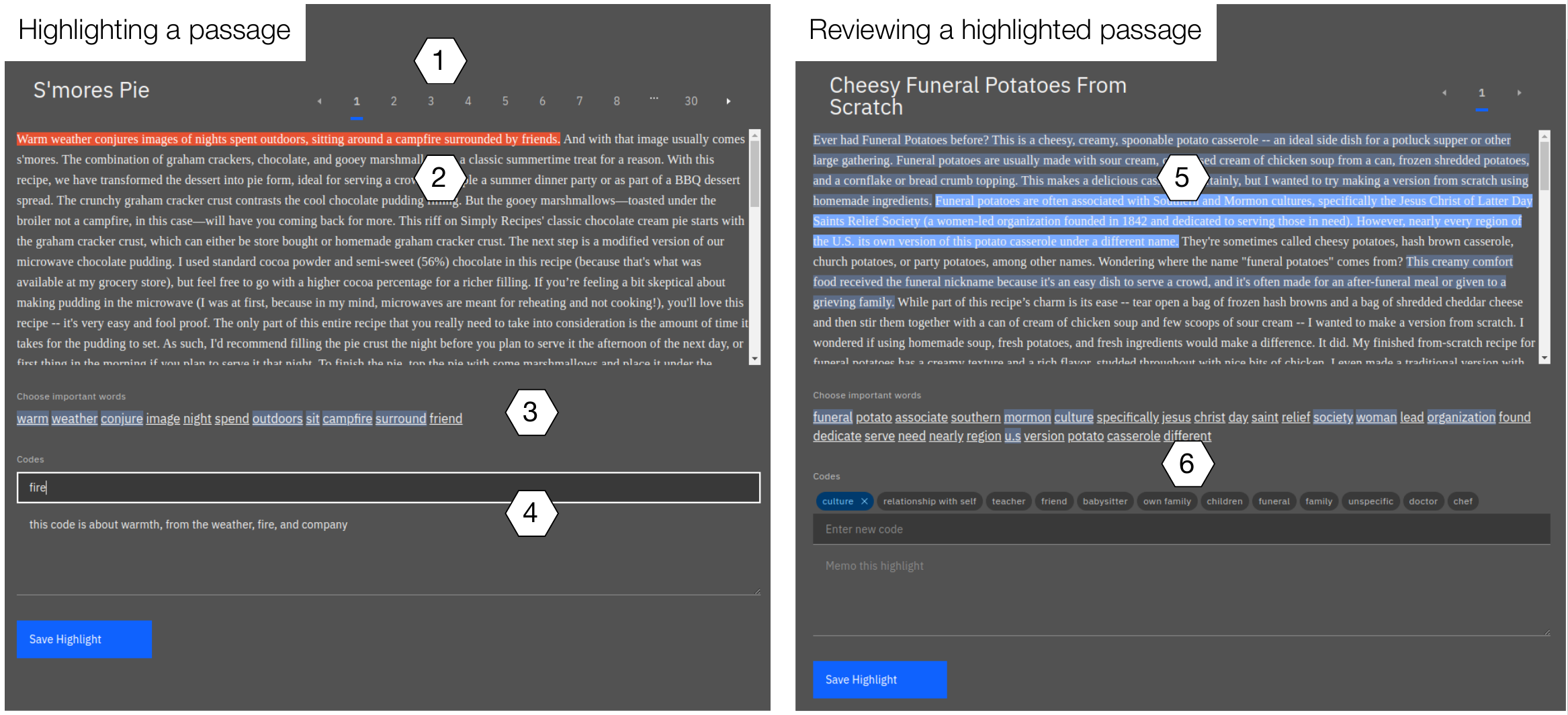}
  \caption{\textit{Scholastic}'s Document Reader features and states. \textmd{When multiple documents are sampled, the analyst can scroll through them using the pagination tool (1). Clicking and dragging across the text creates a highlight (2) that activates the coding footer component (3, 4) that allows the user to apply codes and memos to the highlighted portion of text. The coding footer component also displays a stemmed subset of the highlighted passage (minus stopwords) from which analysts can click to select code-relevant keywords (3). Saving the highlight will cause it to appear as a clickable section of the document text (5). Clicking a previously highlighted section will auto-populate the coding footer component (6) where the analyst can view or edit their previously applied codes, memos, and keywords. }}
  \Description{the document reader is used to apply codes to document passages and select code keywords for the interactive machine learning algorithm.}
  \label{fig:document-reader}
\end{figure*}

\begin{figure*}[h]
  \centering
  \includegraphics[width=\textwidth]{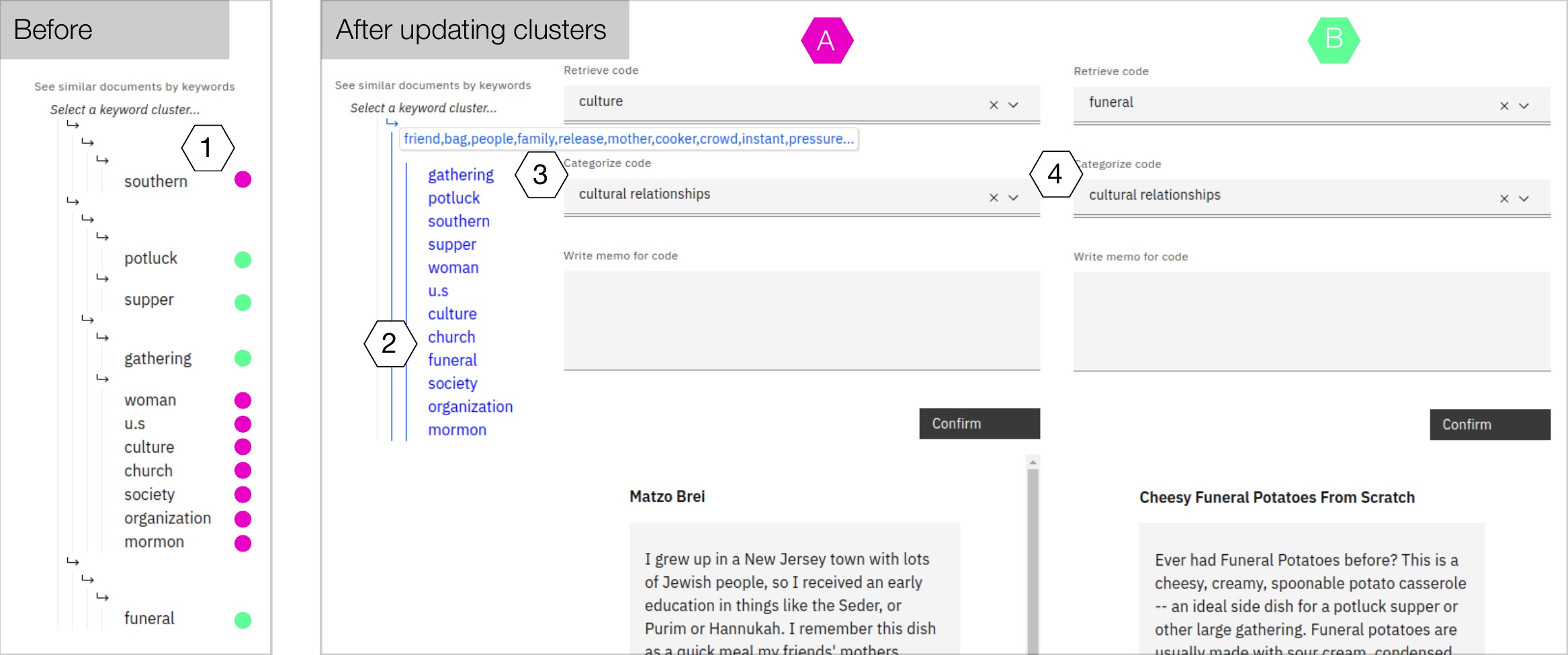}
  \caption{Code Examiner features and states. \textmd{These models show P2's evaluation (\S\ref{sec:task3}) before and after they triggered the interactive model update. The indented tree component (1, 2) displays keywords associated with selected codes, clustered and sorted by word frequency according to the model. Hovering over and selecting a vertical line samples documents most relevant to the word cluster at the given level (2). Note that words corresponding to code A (pink) and code B (green) were not clustered together, and the model update successfully enforced the correct constraints. Moreover, this cluster also found other related words (3) which are shown on hover interaction. These words were clustered together because P2 had categorized codes A and B together (4) using the dropdown menus. \rev{These menus also double as text fields where researchers can modify a code label or create new categories.} Once codes are selected, the highlighted passages are shown at the bottom. These text sections are clickable and will navigate back to the full document within the Document Reader.}}
  \Description{the code examiner is used to categorize codes, and sample more documents using hierarchical word clusters adjacent to already coded documents.}
  \label{fig:word-cluster}
\end{figure*}

To incorporate the above algorithm within a machine-in-the-loop interpretive scholarship workflow, our system prototype \textit{Scholastic} has three views: 
\begin{enumerate}
    \item The Document Map (Figure \ref{fig:document-map}), which supports both random and breadth-first sampling as well as model comparisons;
    \item The Document Reader (Figure \ref{fig:document-reader}), which supports coding, memoing, and keyword selection for the algorithm; and 
    \item The Code Examiner (Figure \ref{fig:word-cluster}), which supports categorization, depth-first search for similar documents related to codes, and subsequent code refinement.
\end{enumerate}
This design embodies qualitative analysts’ workflows for manually applying codes and categories through inductive interpretivist methods while allowing the system to collect metadata (via keyword selection for each applied code) to refine the outputs of our hSBM algorithm. 
The analyst can 
navigate between these views using the navigation bar (Figure \ref{fig:document-map}.1) or by sampling documents. Code filters within the navigation bar filter
overlay pins on the Document Map: pins appear above documents if codes have been applied to passages within those documents. 
The navigation bar also contains a button to update the cluster models on demand.
\rev{The scholar can continue working on their analysis during this model update, which given our study dataset (Appendix \ref{sec:recipes}) and hardware (32GB RAM with an Intel 6-Core i7 processor) took approximately 11 minutes.}
Lastly, the navigation bar also allows a scholar to switch between color themes (light mode by default and dark mode for focused reading) using radio buttons.

\subsection{Breadth-First Sampling: Document Map}

The Document Map serves two main functionalities: corpus exploration and model comparisons. The central visual element is a geographical treemap \cite{Auber2011} representing the hierarchical document clusters as spatial regions.
The analyst controls the granularity of the hierarchical clusters with a step slider (Figure \ref{fig:document-map}.2),
which determines the lowest level cluster boundaries shown. 

Each document is represented with a hexagonal tile (Figure \ref{fig:document-map}.3).
Hovering over individual hexagons on the map dynamically displays the corresponding document title on the preview component at the bottom right. 
Right-clicking a hexagon expands this preview component to show the first 1000 characters of a document without entering the Document Reader; left-clicking a hexagon opens the document in the Document Reader (\S\ref{sec:reader}) to begin coding. 
If the scholar chooses not to use the geographical map for document sampling, 
the Document Map allows them to randomly sample a subset of $N$ documents from the corpus, where the sample size can be specified by the scholar (Figure \ref{fig:document-map}.4). 


We used a geographical treemap 
to draw a familiar visual metaphor between hierarchical document clusters
and maps to allow 
analysts to easily explore the hierarchical clusters. 
\rev{We intentionally do not impose \textit{a priori} cluster keywords or filters on this map to avoid biasing an interpretive scholar's attention toward any specific cluster. We also do not provide information extracted from the model (e.g., common words, topic names, word probabilities) beyond cluster boundaries to avoid bias in cluster interpretation.}
\rev{
By design, the Document Map requires the user to demand details about data items first by hovering over or sampling an item, then build up their own filters and relations between documents (codes and categories), constructing an overview through their own sensemaking process.} 

We display a location pin above a hexagon tile if a code has been applied to the document. 
Each pin is colored according to the category of the applied code. In cases where multiple codes have been applied to a document,
the number of unique categories is initially shown above the text item. 
Analysts can choose to show all codes associated with each document or a subset of codes
using the code filters on the navigation bar (Figure \ref{fig:document-map}.5). 
\rev{The pins are intended to support diverse and adaptive search strategies to explore related (i.e., depth-first) and unrelated (i.e., breadth-first) documents based on the user's ongoing analysis.}
They also allow the scholar to compare their evolving coding schema with the document cluster output.
Once the hSBM has output an updated cluster model on the scholar's demand, they can also make comparisons across the model outputs (using a step slider on the left sidebar) to assess how their inputs affected the distribution of codes across clusters.

\subsection{Applying Codes: Document Reader}
\label{sec:reader}
A scholar accesses the Document Reader (Figure \ref{fig:document-reader}) from either the Document Map (when conducting a breadth-first sampling from the corpus) or the Code Examiner (when revisiting a document given a target code or when conducting a depth-first sampling
using the indented tree; see \S\ref{examiner}). 
When multiple documents have been sampled (e.g., through the indented tree or random sampling), the scholar can scroll through them using the pagination component (Figure \ref{fig:document-reader}.1). 
The Document Reader first displays the selected document’s title and content. The scholar can click-and-drag (i.e., highlight) a passage to apply a code (Figure \ref{fig:document-reader}.2). As soon as the drag is released, the coding footer component appears (Figure \ref{fig:document-reader}.4).

The coding footer component allows the analyst to apply a new code by typing in a code label or apply existing codes by clicking on existing code tags. 
It 
also displays a stemmed subset of words from the highlighted passage (minus stopwords) from which the scholar can choose
relevant keywords for the multilayer hSBM (Figure \ref{fig:document-reader}.3). 
This parsimonious input operation is similar to \textit{in vivo} coding and 
allows the scholar
to characterize semantically meaningful relationships between keywords.
Highlighted passages are then visualized as clickable tagged text \cite{Alexander},
persisting throughout the analysis. 
When revisiting a document, the scholar can click on a previously highlighted passage---which auto-fills the code, keywords, and memos in the coding footer---to allow reflection and refinement (Figure \ref{fig:document-reader}.5, \ref{fig:document-reader}.6).

\subsection{Categorization and Depth-First Search:\\ 
Code Examiner}
\label{examiner}
After applying codes, the Code Examiner (Figure \ref{fig:word-cluster}) allows scholars to compare, refine, and categorize them.
Selecting a pair of codes from drop-down menus populates the interface with each code's label, memos, and the passages containing that code (Figure \ref{fig:word-cluster}.4). 
\rev{Researchers can create a new category for a code by typing in a category label under the code label drop-down. If category labels already exist, this text field also doubles as a drop-down menu to assign an existing category to a code.}
These text segments highlight important keywords selected by the analyst using tagged text \cite{Alexander}. 
Clicking on a passage will navigate the scholar back to the corresponding document in the Document Reader (\S\ref{sec:reader}) for reflection and refinement. 


When codes are selected from the dropdown menus, the Code Examiner displays a pruned tree representation of the hierarchical word clusters generated by the hSBM model (Figure \ref{fig:word-cluster}.2).  
The indented word tree representation of these clusters---similar to ArchiText \cite{Kim2020}---and the simple form-based interface build on familiar metaphors from digital filing systems. 
At the overview level, the tree only displays keywords from the coded passages to reduce complexity, but
hovering over each cluster will display the top ten words within it. 
The analyst can click on a word cluster at various depth levels in the hierarchy to sample 30 documents associated with the selected cluster. 
These 30 documents are selected and displayed in a paginated list (see Figure \ref{fig:document-reader}.1), sorted according to the probability of the selected word cluster occurring in each document $(P(\mathbf{b}^W \mid d))$.
This algorithmic sampling and sorting allows scholars to identify candidate documents depth-first through their code labels and preview the documents in the list view to quickly find those that are most relevant for their current analysis.

\section{Evaluation}
\label{evaluation}

We evaluated \textit{Scholastic} in an interview study with two interpretive researchers. The study consisted of three phases: a brief introduction to the study goals, a think-aloud analysis of 2,615 recipe blog posts using \textit{Scholastic} (Appendix \ref{sec:recipes}), and an exit interview to capture additional feedback. Both participants were trained and published in interpretive research using text data.

\rev{\textit{Scholastic} integrates a constraint-based interactive ML algorithm with both the data sampling and sensemaking loops of qualitative analysis.
Our design considerations took into account that an effective human-AI collaboration tool should support: 1) serendipitous insightful moments during data sampling and sensemaking, and 2) incorporating human input into cluster models (see \S\ref{objectives}).} Therefore, our evaluation sought evidence of following:


\rev{
\begin{itemize}
    \item 
    A range of sampling strategies within the Document Map and the Code Examiner, and
    \item Indications that the interactive ML-related functionalities would not disrupt a scholar’s focused analysis within the Document Reader.
\end{itemize}
}

First, a short introduction was given describing how we wished to examine the ways an interpretive scholar might use our system to analyze a large corpus in order to gather feedback on its design.
We then obtain informed consent to participate and basic demographic information. Finally, we introduced the target dataset and a relevant research question with the following script:

``\textit{You have collected 2615 online recipe blog posts. You are interested in examining how certain foods elicit stories about certain interpersonal relationships as bloggers build narratives around food. Due to the amount of data, you will try a new system designed to support the qualitative analysis of large datasets. The first thing you want to do is to familiarize yourself with your dataset. You open up the system. Talk aloud while you use this screen---as well as the interactions it supports---to examine the dataset.}”

We devised three simple, open-ended tasks to help scaffold the interview such that scholars were able to navigate all system features. 
These task instructions were devised to ensure task coverage, avoid a prescriptive workflow, and minimize researcher bias.
We did not provide participants with tutorials on system functionalities, instead allowing participants to independently learn through their interactions. 
By providing an example dataset, we also intended to mitigate biases that may arise from familiarity with the underlying information.
Each interview took 90 minutes.

\subsection{
Task 1: You open up the system. \\What do you do first?}
Our first task captured initial interactions with the Document Map (\S\ref{interface}.1), which served as a frontispiece for the tool.
P1 described the geographical treemap as ``pleasing,” ``so wholesome,” and that one cluster was ``a very pleasing shape.” They appreciated the connection to familiar geographic maps and board games (e.g. Dungeons \& Dragons, Settlers of Catan). They started exploring documents by hovering over items, which dynamically displayed each item’s title on the right sidebar. Their interactions focused on items at the cluster boundaries, leading them to wonder: ``Is there a reason this cell is bordered off in this section?” 

Although the geographical treemap did not communicate visual content summaries, the dynamic details-on-demand interactions with document titles still allowed P1 to conceptualize clusters in a creative way. They remembered each cluster by its shape and position, developing light-hearted yet memorable names for the regions (e.g., `Grandma’s Cookie Empire’, `Meat-topia’). When adjusting the granularity of the hierarchical document clusters, they talked about how the first and second levels were manageable, but the third level was too granular, saying they might as well be going through the documents manually on their computer. 

Similarly, P2 was drawn to the hover interactions on the geographical treemap, which they described as `techy’ and `aesthetic.’ They noted that although one region seemed to only include pasta recipes, another region seemed to combine drinks and desserts, prompting them to wonder why those recipes were clustered together. P2 described the process of foraging for documents with the map as akin to finding ``little treats.” P2  heavily relied on the right-click preview feature, which retrieves the first 1,000 characters of the document, to make more detailed sense of each item within the clusters. Although P2 did not verbally conceptualize the regions as P1 did, when asked to recall where the pasta and drinks regions were at the end of the interview, P2 located them by their shapes and positions.

\subsection{Task 2: Sample document items \\and apply codes.}
How participants sampled documents were left to their preferences. P1 chose to select documents via click interactions on the map, whereas P2 sampled ten random documents. With the Document Reader (\S\ref{interface}.3) shown, participants were then instructed to begin applying codes. P1’s coding process started with `meme’ codes (e.g., `veggie tales,’ `boil em mash em’) that they later refined. Neither participants had guidance for why the keyword selection feature was present, but immediately speculated that keywords might be used to help update the cluster outputs, as P1 noted: ``The machine will operate better if I give it more input, is what I'm assuming there.” P2 also recognized that the keywords could benefit collaboration with the machine: ``so when we see this in other documents, [the system is] going to assign more weight to that in some way.” To this point, they speculated about what would happen if they trained the model incorrectly: ``I feel like it's basically the garbage in, garbage out principle.” 

P1 appreciated that the keyword selection features in the Document Reader were automatically stemmed and stripped of stopwords, but noted that stopwords may be useful for in-depth sociolinguistic analysis. For P2, on the other hand, keyword selection became a redundant feature because they began by coding short chunks of texts with one to three words (\textit{in vivo} coding). However, when they started highlighting larger passages, they commented how selecting keywords within these chunks, in combination with memoing, could potentially aid in their self-reflection on the meaning of a highlighted passage. Lastly, both experts noticed a few missing features for coding: notably, the ability to apply multiple codes to a single highlighted passage or to simultaneously highlight two passages and apply a single code to both.

\subsection{Task 3: Now that you have codes, \\organize them into categories.}
\label{sec:task3}
Upon being given this instruction, participants intuitively navigated to the Code Examiner (\S\ref{interface}.3) using the navigation bar to reflect and categorize the codes they applied in the Document Reader. On the Code Examiner, P1 selected a code (`meat-lovers’) and saw the hierarchical word clusters, pruned to include only clusters with associated code keywords. Clicking into one of the clusters prompted the sampling of 30 documents, ordered by the likelihood of containing the clicked word cluster. P1 could not immediately make sense of why these documents were sampled, since some documents did not contain the keyword, instead containing other words related to them based on the clustering output. Also, interesting passages within recipe blog posts may be sparsely scattered, only serving as a transition from the introduction to the recipe body. However, their reaction then was to code ``more of this [document] with like, people's declaration of love to meat,” since they developed an understanding that the document was sampled due to its relationship to their keywords from the combined visualizations and raw text.

When P2 started categorizing codes, they noticed that the colored code tags in the navigation bar (Figure \ref{fig:document-map}.5) made it easy to see how the categories were emerging as well as which codes were uncategorized. In exploring the word cluster list, P2 clicked into a leaf cluster containing the keyword `friend.’ They used Ctrl-F to find occurrences of the word “friend” from the word cluster, which was present in the second document but not the first. They noted that the cluster levels seemed to get ``more filtered” down the hierarchy. For example, they noticed the associated words at the lower two cluster levels started with the word `friend,’ (Figure \ref{fig:word-cluster}.3) but words at the top level cluster began with `salt.’ They then felt more certain about the function of the indented tree when working with their ``culture” code: ``Let's see what's in the Southern cluster. So on the first level, it's `Southern,’ `biscuits,’ `health’; on the second level, it's `cake,’ `Southern’; and on the third, it's `roll,’ `dough,’ `filling,’ etc. Again, I do feel this is grabbing all the documents that have `Southern.’ or maybe something like that.”

Due to time constraints, only P2 was able to explore the document and word clusters updated based on their codes and categories. After the model update (which took around 11 minutes), they interacted with the Document Map with the color-categorized code filters, doing this with several codes to see where the coded documents now appeared. They contrasted the document cluster partitions with the positions of colored code pins, discussing how it was interesting that documents with the same codes still appeared across different clusters. They had expected that documents sharing codes would instead be clustered together by the update. Upon being asked what they would do next, they responded that they would keep coding to see if more concrete patterns might emerge when contrasting the document cluster partitions with the positions of colored code pins.

\section{Discussion}
\label{discussion}
\textit{Scholastic}
is a human-AI collaboration tool for interpretive scholarship co-designed with experts. The system 
represents preliminary steps towards the vision of supporting scalable
qualitative analysis 
with large text corpora scaffolded by algorithmic and visualization tools. We took a human-centered approach to enabling this epistemic practice, grounding our design objectives in our team’s earlier user interviews \cite{Jiang2021, Feuston2021} and design iterations between the interpretive scholars and visualization researchers on our team. 
Our discussions revealed the importance of designing for agency: tools should enhance analysts' natural workflows rather than enforcing alternative practices. On the algorithmic side, models should adapt to human input from the scholar's analysis; on the visualization side, the visualization should evolve to incorporate human-inferred codes and categories. Here, we summarize preliminary outcomes from the implementation and evaluation of \textit{Scholastic} to inform future work on forging effective human-AI collaboration for inductive and interpretive text analysis.

\subsection{Outcome 1. Supporting Serendipity}
\rev{Most cluster visualizations follow the visual information-seeking mantra, starting with descriptive visual summaries of the data (overviews) that are interactively adjusted (filter, relate), and information about individual data points is available on-demand. The goal of these techniques is to better provide quantitative summaries of cluster contents at-a-glance. In contrast, our geographical treemap communicated only the size and hierarchical containment of each cluster: \textit{Scholastic} does not impose \textit{a priori} cluster keywords or filters to avoid biasing an interpretive and inductive researcher’s attention toward any specific cluster. By design, the Document Map requires the user to demand details first, then build up their own filters and relations (i.e., codes and categories), constructing an overview through their own sensemaking process. This reversal of the information-seeking mantra allows the analyst to either implicitly develop their own mental model of a cluster’s meaning or explicitly code documents and allow the interactive machine learning algorithm to match their outputs more closely to codes and categories identified by analysts.}

However, the analyst’s 
curiosity about emergent features on the geographical treemap became an entry point for interpretation. Their use and interpretation of these 
features naturally shifted toward data sensemaking.
They appeared to be able to integrate both text and shape to construct a better mental model of the space of documents \cite{tversky1991spatial},
to efficiently sample data items,
avoiding data items from the same cluster or sampling data items from a cluster of interest.
For P1, naming regions (e.g., ``meatopia”) provided a way to remember what documents had been sampled and to revisit similar documents later (e.g., ``maybe I should go look at what's in Meatopia”). They called the regions by names related to their shape or the document content: ``And then there's this little cell here, that's like the country on the African continent, that's like the little donut hole…”). Although P2 initially expressed confusion over some document clusters, their ability to vocalize their uncertainty (e.g., ``why does [this cluster] contain both drinks and desserts?'') also reflected curiosity-driven exploration by both verbal and spatial conceptualization of the map \cite{Dickson1988}.

\subsection{Outcome 2. Right Place, Right Time}
By 
incorporating evolving human codes and categories into \textit{Scholastic}'s interfaces and algorithms, we supported both random and strategic sampling with visualizations that increasingly reflect the knowledge built by the user rather than by the raw text models.
The participants in our evaluation each chose a different strategy for sampling. 
Supporting diverse strategies gives the researchers the agency to choose the right tools for the right data, scenarios, and times. Although P2 chose not to use the map for sampling initially, in our post-study interview they appreciated that the document clusters helped them familiarize themselves with the breadth of the dataset before 
coding. 

P2 noticed that keyword selection for the interactive ML algorithm could help them analyze a passage in more depth.
They mentioned that coding individual keywords would be ``the kind of thing I would probably be memoing about,'' as identifying keywords within codes allows analysts to focus on \textit{why} a particular code might be appropriate for a given passage.
In this sense, both participants 
saw the tool as a collaborator, where the ML features helped
the analyst, but the analyst 
also helped the model evolve. P1 in many ways personified the system as they would a research assistant. They remarked when there was a lag in saving a highlight that ``He's keeping up in the back there.'' P1 noted that if the keyword selection appeared beneficial for the model, they would be pleased “because it's designed to help me. It is my helper.” P2 noted that seeing how the document cluster outputs had changed based on input made the model more trustworthy, because it was able to adapt to their own interpretations. 

\section{Limitations and Future Work}
\label{limitations}
\textit{Scholastic}'s design focused on capturing the core components of qualitative analysis workflows without necessarily supporting any individual methodology (e.g., grounded theory or thematic analysis). In our evaluation with qualitative experts, analysts wanted several additional features to 
tailor the tool for specific interpretive approaches. For example, P1 appreciated that the keyword selection features in the Document Reader were automatically stemmed and stripped of stopwords within the context of their preferred methods (thematic analysis). However, both participants noted that stopwords may be useful for in-depth sociolinguistic analyses. Future work should explore extensible frameworks that tailor analysis support to individual methodological or disciplinary needs. 

We note that our tool 
does not aim to incorporate all of the coding features present in commercial tools like MaxQDA that primarily support code management. 
Analysts’ coding practices vary widely: for example, P1 would have preferred to code titles of documents in addition to their body texts. Analysts may desire to highlight multiple disjoint segments with a single code or to assign multiple unique codes to a single highlight. 
The latter practice---while frequently requested---requires further algorithmic development since our algorithm assumes that each keyword maps to a single code. If a user wishes to apply multiple codes to the same passage of text in our current implementation, the text must be segmented into unique keywords for each code to avoid overlapping assignments.
For interpretive scholarship, assigning multiple codes to a single passage for approaches 
treating text data as a bag-of-words input will require novel algorithmic support, unless the word tokens can be separated according to additional metadata such as semantics and syntax as in Griffiths et al. \cite{Griffiths2005}. 

Engagement played a significant role in analysts' desire to use a given visualization for sampling data items. For example, P1 noted that the directory-like nature of the indented tree visualization
lacked the engaging, more organic features that had emerged on the
Document Map. This difference impacted how willing they were to sample documents with the indented tree visualization. 
\rev{Our future work will study how visual features emerging in cluster visualizations may play a role in learning and memory for qualitative data.}
\rev{For example, even in the absence of explicit visual summaries, people were able to leverage details-on-demand interactions and emergent shapes of the geographical treemap to conceptualize and remember the information contained within clusters and guide their exploration. 
Future work should explore if this behavior has a capacity limit (e.g., number of clusters) or is mediated by the visualization technique used (e.g., geographical treemaps vs. scatterplots).}

\rev{Our evaluation studies focused on the system's usability with two researchers. An extended, longitudinal evaluation could better demonstrate analytical insights generated using \textit{Scholastic}. We intend to deploy this system as part of a future longitudinal study on the impact of mixed-initiative tools on qualitative analysis outcomes.
This comparative study may investigate the varying efficiency and efficacy of these tools over a lengthy collaborative and interpretive text analysis (e.g., by analyzing subjective evaluations of confidence and trust in analysts' knowledge work and comparing research outcomes across users).}


\section{Conclusion}
\label{conclusion}
Statistical models are powerful tools for analyzing text data. However, there is no consensus within the interpretivist research community regarding what the role of machine learning should be within their practices \cite{dimaggio2013exploiting, Nelson2020}.
Our co-design process with experts took a human-centered approach to implement an interface and algorithm for supporting key phases of interpretive and inductive text analysis workflows \cite{Baumer2017a}. 
\textit{Scholastic} embodies the goals of designing for various sampling strategies given a corpus, letting the AI model adapt to on-going knowledge development, and allowing human sensemaking to drive interpretive text analysis, enabling familiar and non-disruptive interactions with the AI mediated by visualizations. Given the popularity of qualitative methods for analyzing text data within human-computer interaction \cite{McDonald2019} and visual analytics \cite{Diehl2022}, we hope that our work will build a foundation for future mixed-initiative systems scaffolding interpretive scholarship.

\begin{acks}
 The authors would like to thank Kandrea Wade and Casey Fiesler for their input at the conceptualization stages, students and faculty at the ATLAS Institute, CU Boulder during the design, development, and writing stages, and Matteo Abrate (https://bl.ocks.org/nitaku) for his valuable examples of geographical treemaps. This work was supported by NSF awards \#1764092 \& \#2046725.
\end{acks}

\balance
\bibliographystyle{ACM-Reference-Format}
\bibliography{qbd}


\begin{thebibliography}{50}


\ifx \showCODEN    \undefined \def \showCODEN     #1{\unskip}     \fi
\ifx \showDOI      \undefined \def \showDOI       #1{#1}\fi
\ifx \showISBNx    \undefined \def \showISBNx     #1{\unskip}     \fi
\ifx \showISBNxiii \undefined \def \showISBNxiii  #1{\unskip}     \fi
\ifx \showISSN     \undefined \def \showISSN      #1{\unskip}     \fi
\ifx \showLCCN     \undefined \def \showLCCN      #1{\unskip}     \fi
\ifx \shownote     \undefined \def \shownote      #1{#1}          \fi
\ifx \showarticletitle \undefined \def \showarticletitle #1{#1}   \fi
\ifx \showURL      \undefined \def \showURL       {\relax}        \fi
\providecommand\bibfield[2]{#2}
\providecommand\bibinfo[2]{#2}
\providecommand\natexlab[1]{#1}
\providecommand\showeprint[2][]{arXiv:#2}

\bibitem[Alexander et~al\mbox{.}(2015)]%
        {Alexander}
\bibfield{author}{\bibinfo{person}{Eric Alexander}, \bibinfo{person}{Joe
  Kohlmann}, \bibinfo{person}{Robin Valenza}, \bibinfo{person}{Michael
  Witmore}, {and} \bibinfo{person}{Michael Gleicher}.}
  \bibinfo{year}{2015}\natexlab{}.
\newblock \showarticletitle{{Serendip: Topic model-driven visual exploration of
  text corpora}}. In \bibinfo{booktitle}{\emph{2014 IEEE Conference on Visual
  Analytics Science and Technology, VAST 2014 - Proceedings}}.
  \bibinfo{pages}{173--182}.
\newblock
\showISBNx{9781479962273}
\urldef\tempurl%
\url{https://doi.org/10.1109/VAST.2014.7042493}
\showDOI{\tempurl}


\bibitem[Auber et~al\mbox{.}(2011)]%
        {Auber2011}
\bibfield{author}{\bibinfo{person}{D Auber}, \bibinfo{person}{C Huet},
  \bibinfo{person}{A Lambert}, {and} \bibinfo{person}{A Sallaberry}.}
  \bibinfo{year}{2011}\natexlab{}.
\newblock \showarticletitle{{Geographical treemaps}}.
\newblock \bibinfo{journal}{\emph{Not yet published}} \bibinfo{number}{March}
  (\bibinfo{year}{2011}).
\newblock
\urldef\tempurl%
\url{http://www.labri.fr/perso/auber/download/paper{\_}TVCG/docPourTVCG/InfoVisVersion.pdf}
\showURL{%
\tempurl}


\bibitem[Baumer(2017)]%
        {Baumer2017a}
\bibfield{author}{\bibinfo{person}{Eric P~S Baumer}.}
  \bibinfo{year}{2017}\natexlab{}.
\newblock \showarticletitle{{Toward human-centered algorithm design}}.
\newblock \bibinfo{journal}{\emph{Big Data {\&} Society}} \bibinfo{volume}{4},
  \bibinfo{number}{2} (\bibinfo{year}{2017}), \bibinfo{pages}{205395171771885}.
\newblock
\showISSN{2053-9517}
\urldef\tempurl%
\url{https://doi.org/10.1177/2053951717718854}
\showDOI{\tempurl}


\bibitem[Baumer et~al\mbox{.}(2017)]%
        {Baumer2017}
\bibfield{author}{\bibinfo{person}{Eric P~S Baumer}, \bibinfo{person}{David
  Mimno}, \bibinfo{person}{Shion Guha}, \bibinfo{person}{Emily Quan}, {and}
  \bibinfo{person}{Geri~K Gay}.} \bibinfo{year}{2017}\natexlab{}.
\newblock \showarticletitle{{Comparing grounded theory and topic modeling:
  Extreme divergence or unlikely convergence?}}
\newblock \bibinfo{journal}{\emph{Journal of the Association for Information
  Science and Technology}} \bibinfo{volume}{68}, \bibinfo{number}{6}
  (\bibinfo{year}{2017}), \bibinfo{pages}{1397--1410}.
\newblock
\showISSN{2330-1635}
\urldef\tempurl%
\url{https://doi.org/10.1002/asi.23786}
\showDOI{\tempurl}


\bibitem[Baumer et~al\mbox{.}(2020)]%
        {Baumer2020a}
\bibfield{author}{\bibinfo{person}{Eric P~S Baumer}, \bibinfo{person}{Drew
  Siedel}, \bibinfo{person}{Lena Mcdonnell}, \bibinfo{person}{Jiayun Zhong},
  \bibinfo{person}{Patricia Sittikul}, {and} \bibinfo{person}{Micki Mcgee}.}
  \bibinfo{year}{2020}\natexlab{}.
\newblock \showarticletitle{{Topicalizer: reframing core concepts in machine
  learning visualization by co-designing for interpretivist scholarship}}.
\newblock \bibinfo{journal}{\emph{Human–Computer Interaction}}
  (\bibinfo{year}{2020}), \bibinfo{pages}{1--29}.
\newblock
\showISSN{0737-0024}
\urldef\tempurl%
\url{https://doi.org/10.1080/07370024.2020.1734460}
\showDOI{\tempurl}


\bibitem[Blei et~al\mbox{.}(2003)]%
        {Blei2003}
\bibfield{author}{\bibinfo{person}{David~M Blei}, \bibinfo{person}{Andrew~Y
  Ng}, {and} \bibinfo{person}{Michael~I Jordan}.}
  \bibinfo{year}{2003}\natexlab{}.
\newblock \bibinfo{booktitle}{\emph{{Latent Dirichlet Allocation}}}.
\newblock \bibinfo{type}{{T}echnical {R}eport}. \bibinfo{pages}{993--1022}
  pages.
\newblock


\bibitem[Boyd-Graber et~al\mbox{.}(2017)]%
        {Boyd-Graber2017}
\bibfield{author}{\bibinfo{person}{Jordan Boyd-Graber},
  \bibinfo{person}{Yuening Hu}, {and} \bibinfo{person}{David Mimno}.}
  \bibinfo{year}{2017}\natexlab{}.
\newblock \showarticletitle{{Applications of Topic Models}}.
\newblock \bibinfo{journal}{\emph{Applications of Topic Models}}
  \bibinfo{volume}{XX}, \bibinfo{number}{Xx} (\bibinfo{year}{2017}),
  \bibinfo{pages}{1--154}.
\newblock
\urldef\tempurl%
\url{https://doi.org/10.1561/9781680833096}
\showDOI{\tempurl}


\bibitem[Braun and Clarke(2006)]%
        {Braun2006}
\bibfield{author}{\bibinfo{person}{Virginia Braun} {and}
  \bibinfo{person}{Victoria Clarke}.} \bibinfo{year}{2006}\natexlab{}.
\newblock \showarticletitle{{Using thematic analysis in psychology}}.
\newblock \bibinfo{journal}{\emph{Qualitative Research in Psychology}}
  \bibinfo{volume}{3}, \bibinfo{number}{2} (\bibinfo{year}{2006}),
  \bibinfo{pages}{77--101}.
\newblock
\showISSN{14780887}
\urldef\tempurl%
\url{https://doi.org/10.1191/1478088706qp063oa}
\showDOI{\tempurl}


\bibitem[Brehmer et~al\mbox{.}(2014)]%
        {Brehmer2014}
\bibfield{author}{\bibinfo{person}{Matthew Brehmer}, \bibinfo{person}{Stephen
  Ingram}, \bibinfo{person}{Jonathan Stray}, {and} \bibinfo{person}{Tamara
  Munzner}.} \bibinfo{year}{2014}\natexlab{}.
\newblock \showarticletitle{{Overview: The Design, Adoption, and Analysis of a
  Visual Document Mining Tool for Investigative Journalists}}.
\newblock \bibinfo{journal}{\emph{IEEE Transactions on Visualization and
  Computer Graphics}} \bibinfo{volume}{20}, \bibinfo{number}{12}
  (\bibinfo{year}{2014}), \bibinfo{pages}{2271--2280}.
\newblock
\showISSN{1077-2626}
\urldef\tempurl%
\url{https://doi.org/10.1109/tvcg.2014.2346431}
\showDOI{\tempurl}


\bibitem[Chandrasegaran et~al\mbox{.}(2017)]%
        {Chandrasegaran2017c}
\bibfield{author}{\bibinfo{person}{Senthil Chandrasegaran},
  \bibinfo{person}{Sriram~Karthik Badam}, \bibinfo{person}{Lorraine
  Kisselburgh}, \bibinfo{person}{Karthik Ramani}, {and} \bibinfo{person}{Niklas
  Elmqvist}.} \bibinfo{year}{2017}\natexlab{}.
\newblock \showarticletitle{{Integrating Visual Analytics Support for Grounded
  Theory Practice in Qualitative Text Analysis}}.
\newblock \bibinfo{journal}{\emph{Computer Graphics Forum}}
  \bibinfo{volume}{36}, \bibinfo{number}{3} (\bibinfo{date}{jun}
  \bibinfo{year}{2017}), \bibinfo{pages}{201--212}.
\newblock
\showISSN{14678659}
\urldef\tempurl%
\url{https://doi.org/10.1111/cgf.13180}
\showDOI{\tempurl}


\bibitem[Chen et~al\mbox{.}(2018)]%
        {Chen2018a}
\bibfield{author}{\bibinfo{person}{Nan-Chen Chen}, \bibinfo{person}{Margaret
  Drouhard}, \bibinfo{person}{Rafal Kocielnik}, \bibinfo{person}{Jina Suh},
  {and} \bibinfo{person}{Cecilia~R Aragon}.} \bibinfo{year}{2018}\natexlab{}.
\newblock \showarticletitle{{Using Machine Learning to Support Qualitative
  Coding in Social Science}}.
\newblock \bibinfo{journal}{\emph{ACM Transactions on Interactive Intelligent
  Systems}} \bibinfo{volume}{8}, \bibinfo{number}{2} (\bibinfo{year}{2018}),
  \bibinfo{pages}{1--20}.
\newblock
\showISSN{2160-6455}
\urldef\tempurl%
\url{https://doi.org/10.1145/3185515}
\showDOI{\tempurl}


\bibitem[Choo et~al\mbox{.}(2013)]%
        {Choo2013}
\bibfield{author}{\bibinfo{person}{Jaegul Choo}, \bibinfo{person}{Changhyun
  Lee}, \bibinfo{person}{Chandan~K Reddy}, {and} \bibinfo{person}{Haesun
  Park}.} \bibinfo{year}{2013}\natexlab{}.
\newblock \showarticletitle{{UTOPIAN: User-Driven Topic Modeling Based on
  Interactive Nonnegative Matrix Factorization}}.
\newblock \bibinfo{journal}{\emph{IEEE Transactions on Visualization and
  Computer Graphics}} \bibinfo{volume}{19}, \bibinfo{number}{12}
  (\bibinfo{year}{2013}), \bibinfo{pages}{1992--2001}.
\newblock
\showISSN{1077-2626}
\urldef\tempurl%
\url{https://doi.org/10.1109/tvcg.2013.212}
\showDOI{\tempurl}


\bibitem[Chuang et~al\mbox{.}(2012)]%
        {Chuang2012}
\bibfield{author}{\bibinfo{person}{Jason Chuang},
  \bibinfo{person}{Christopher~D. Manning}, {and} \bibinfo{person}{Jeffrey
  Heer}.} \bibinfo{year}{2012}\natexlab{}.
\newblock \showarticletitle{{Termite: Visualization techniques for assessing
  textual topic models}}.
\newblock \bibinfo{journal}{\emph{Proceedings of the Workshop on Advanced
  Visual Interfaces AVI}} (\bibinfo{year}{2012}), \bibinfo{pages}{74--77}.
\newblock
\showISBNx{9781450312875}
\urldef\tempurl%
\url{https://doi.org/10.1145/2254556.2254572}
\showDOI{\tempurl}


\bibitem[Chuang et~al\mbox{.}(2015)]%
        {Chuang2015}
\bibfield{author}{\bibinfo{person}{Jason Chuang}, \bibinfo{person}{John~D
  Wilkerson}, \bibinfo{person}{Brandon~M Stewart}, {and}
  \bibinfo{person}{Margaret~E Roberts}.} \bibinfo{year}{2015}\natexlab{}.
\newblock \showarticletitle{{Computer-Assisted Content Analysis : Topic Models
  for Exploring Multiple Subjective Interpretations}}.
\newblock \bibinfo{journal}{\emph{NIPS Workshop on Human-Propelled Machine
  Learnin}} (\bibinfo{year}{2015}), \bibinfo{pages}{1--9}.
\newblock


\bibitem[Collier and Mahoney(1996)]%
        {collier_mahoney_1996}
\bibfield{author}{\bibinfo{person}{David Collier} {and} \bibinfo{person}{James
  Mahoney}.} \bibinfo{year}{1996}\natexlab{}.
\newblock \showarticletitle{Insights and Pitfalls: Selection Bias in
  Qualitative Research}.
\newblock \bibinfo{journal}{\emph{World Politics}} \bibinfo{volume}{49},
  \bibinfo{number}{1} (\bibinfo{year}{1996}), \bibinfo{pages}{56–91}.
\newblock
\urldef\tempurl%
\url{https://doi.org/10.1353/wp.1996.0023}
\showDOI{\tempurl}


\bibitem[Corbin and Strauss(1990)]%
        {Corbin1990}
\bibfield{author}{\bibinfo{person}{Juliet~M. Corbin} {and}
  \bibinfo{person}{Anselm Strauss}.} \bibinfo{year}{1990}\natexlab{}.
\newblock \showarticletitle{{Grounded theory research: Procedures, canons, and
  evaluative criteria}}.
\newblock \bibinfo{journal}{\emph{Qualitative Sociology}} \bibinfo{volume}{13},
  \bibinfo{number}{1} (\bibinfo{year}{1990}), \bibinfo{pages}{3--21}.
\newblock
\showISSN{01620436}
\urldef\tempurl%
\url{https://doi.org/10.1007/BF00988593}
\showDOI{\tempurl}


\bibitem[Dickson et~al\mbox{.}(1988)]%
        {Dickson1988}
\bibfield{author}{\bibinfo{person}{Lou Ann~S. Dickson},
  \bibinfo{person}{Philipp~S. Schrankel}, {and} \bibinfo{person}{Raymond~W.
  Kulhavy}.} \bibinfo{year}{1988}\natexlab{}.
\newblock \showarticletitle{{Verbal and spatial encoding of text}}.
\newblock \bibinfo{journal}{\emph{Instructional Science}} \bibinfo{volume}{17},
  \bibinfo{number}{2} (\bibinfo{year}{1988}), \bibinfo{pages}{145--157}.
\newblock
\showISSN{00204277}
\urldef\tempurl%
\url{https://doi.org/10.1007/BF00052700}
\showDOI{\tempurl}


\bibitem[Diehl et~al\mbox{.}(2022)]%
        {Diehl2022}
\bibfield{author}{\bibinfo{person}{Alexandra Diehl}, \bibinfo{person}{Alfie
  Abdul-Rahman}, \bibinfo{person}{Benjamin Bach}, \bibinfo{person}{Mennatallah
  El-Assady}, \bibinfo{person}{Matthias Kraus}, \bibinfo{person}{Robert~S.
  Laramee}, {and} \bibinfo{person}{Min Chen}.} \bibinfo{year}{2022}\natexlab{}.
\newblock \showarticletitle{{Characterizing Grounded Theory Approaches in
  Visualization}}.
\newblock  \bibinfo{volume}{41}, \bibinfo{number}{3} (\bibinfo{year}{2022}).
\newblock
\showeprint[arxiv]{2203.01777}
\urldef\tempurl%
\url{http://arxiv.org/abs/2203.01777}
\showURL{%
\tempurl}


\bibitem[DiMaggio et~al\mbox{.}(2013)]%
        {dimaggio2013exploiting}
\bibfield{author}{\bibinfo{person}{Paul DiMaggio}, \bibinfo{person}{Manish
  Nag}, {and} \bibinfo{person}{David Blei}.} \bibinfo{year}{2013}\natexlab{}.
\newblock \showarticletitle{Exploiting affinities between topic modeling and
  the sociological perspective on culture: Application to newspaper coverage of
  US government arts funding}.
\newblock \bibinfo{journal}{\emph{Poetics}} \bibinfo{volume}{41},
  \bibinfo{number}{6} (\bibinfo{year}{2013}), \bibinfo{pages}{570--606}.
\newblock


\bibitem[Drouhard et~al\mbox{.}(2017)]%
        {Drouhard2017a}
\bibfield{author}{\bibinfo{person}{Margaret Drouhard},
  \bibinfo{person}{Nan-Chen Chen}, \bibinfo{person}{Jina Suh},
  \bibinfo{person}{Rafal Kocielnik}, \bibinfo{person}{Vanessa Pena-Araya},
  \bibinfo{person}{Keting Cen}, \bibinfo{person}{Xiangyi Zheng}, {and}
  \bibinfo{person}{Cecilia~R Aragon}.} \bibinfo{year}{2017}\natexlab{}.
\newblock \showarticletitle{{Aeonium: Visual analytics to support collaborative
  qualitative coding}}. In \bibinfo{booktitle}{\emph{2017 IEEE Pacific
  Visualization Symposium (PacificVis)}}. \bibinfo{publisher}{IEEE}.
\newblock
\urldef\tempurl%
\url{https://doi.org/10.1109/pacificvis.2017.8031598}
\showDOI{\tempurl}


\bibitem[El-Assady et~al\mbox{.}(2020)]%
        {El-Assady2020}
\bibfield{author}{\bibinfo{person}{Mennatallah El-Assady},
  \bibinfo{person}{Rebecca Kehlbeck}, \bibinfo{person}{Christopher Collins},
  \bibinfo{person}{Daniel Keim}, {and} \bibinfo{person}{Oliver Deussen}.}
  \bibinfo{year}{2020}\natexlab{}.
\newblock \showarticletitle{{Semantic concept spaces: Guided topic model
  refinement using word-embedding projections}}.
\newblock \bibinfo{journal}{\emph{IEEE Transactions on Visualization and
  Computer Graphics}} \bibinfo{volume}{26}, \bibinfo{number}{1}
  (\bibinfo{year}{2020}), \bibinfo{pages}{1001--1011}.
\newblock
\showISSN{19410506}
\urldef\tempurl%
\url{https://doi.org/10.1109/TVCG.2019.2934654}
\showDOI{\tempurl}
\showeprint[arxiv]{1908.00475}


\bibitem[Endert et~al\mbox{.}(2012)]%
        {Endert2012}
\bibfield{author}{\bibinfo{person}{Alex Endert}, \bibinfo{person}{Patrick
  Fiaux}, {and} \bibinfo{person}{Chris North}.}
  \bibinfo{year}{2012}\natexlab{}.
\newblock \showarticletitle{{Semantic interaction for sensemaking: Inferring
  analytical reasoning for model steering}}.
\newblock \bibinfo{journal}{\emph{IEEE Transactions on Visualization and
  Computer Graphics}} \bibinfo{volume}{18}, \bibinfo{number}{12}
  (\bibinfo{year}{2012}), \bibinfo{pages}{2879--2889}.
\newblock
\showISBNx{10772626/12}
\showISSN{10772626}
\urldef\tempurl%
\url{https://doi.org/10.1109/TVCG.2012.260}
\showDOI{\tempurl}


\bibitem[Feuston and Brubaker(2021)]%
        {Feuston2021}
\bibfield{author}{\bibinfo{person}{Jessica~L. Feuston} {and}
  \bibinfo{person}{Jed~R. Brubaker}.} \bibinfo{year}{2021}\natexlab{}.
\newblock \showarticletitle{{Putting Tools in Their Place: The Role of Time and
  Perspective in Human-AI Collaboration for Qualitative Analysis}}.
\newblock \bibinfo{journal}{\emph{Proceedings of the ACM on Human-Computer
  Interaction}} \bibinfo{volume}{5}, \bibinfo{number}{CSCW2}
  (\bibinfo{year}{2021}).
\newblock
\showISSN{25730142}
\urldef\tempurl%
\url{https://doi.org/10.1145/3479856}
\showDOI{\tempurl}


\bibitem[Gauthier and Wallace(2022)]%
        {Gauthier2022}
\bibfield{author}{\bibinfo{person}{Robert~P. Gauthier} {and}
  \bibinfo{person}{James~R. Wallace}.} \bibinfo{year}{2022}\natexlab{}.
\newblock \showarticletitle{{The Computational Thematic Analysis Toolkit}}.
\newblock \bibinfo{journal}{\emph{Proceedings of the ACM on Human-Computer
  Interaction}} \bibinfo{volume}{6}, \bibinfo{number}{GROUP}
  (\bibinfo{year}{2022}), \bibinfo{pages}{1--15}.
\newblock
\showISSN{25730142}
\urldef\tempurl%
\url{https://doi.org/10.1145/3492844}
\showDOI{\tempurl}


\bibitem[Gerlach et~al\mbox{.}(2018)]%
        {Gerlach2018}
\bibfield{author}{\bibinfo{person}{Martin Gerlach}, \bibinfo{person}{Tiago~P
  Peixoto}, {and} \bibinfo{person}{Eduardo~G Altmann}.}
  \bibinfo{year}{2018}\natexlab{}.
\newblock \showarticletitle{{A network approach to topic models}}.
\newblock \bibinfo{journal}{\emph{Science Advances}} \bibinfo{volume}{4},
  \bibinfo{number}{7} (\bibinfo{year}{2018}), \bibinfo{pages}{eaaq1360}.
\newblock
\showISSN{2375-2548}
\urldef\tempurl%
\url{https://doi.org/10.1126/sciadv.aaq1360}
\showDOI{\tempurl}


\bibitem[Green and Chen(2019a)]%
        {green2019disparate}
\bibfield{author}{\bibinfo{person}{Ben Green} {and} \bibinfo{person}{Yiling
  Chen}.} \bibinfo{year}{2019}\natexlab{a}.
\newblock \showarticletitle{Disparate interactions: An algorithm-in-the-loop
  analysis of fairness in risk assessments}. In
  \bibinfo{booktitle}{\emph{Proceedings of the conference on fairness,
  accountability, and transparency}}. \bibinfo{pages}{90--99}.
\newblock


\bibitem[Green and Chen(2019b)]%
        {green2019principles}
\bibfield{author}{\bibinfo{person}{Ben Green} {and} \bibinfo{person}{Yiling
  Chen}.} \bibinfo{year}{2019}\natexlab{b}.
\newblock \showarticletitle{The principles and limits of algorithm-in-the-loop
  decision making}.
\newblock \bibinfo{journal}{\emph{Proceedings of the ACM on Human-Computer
  Interaction}} \bibinfo{volume}{3}, \bibinfo{number}{CSCW}
  (\bibinfo{year}{2019}), \bibinfo{pages}{1--24}.
\newblock


\bibitem[Griffiths et~al\mbox{.}(2005)]%
        {Griffiths2005}
\bibfield{author}{\bibinfo{person}{Thomas~L. Griffiths}, \bibinfo{person}{Mark
  Steyvers}, \bibinfo{person}{David~M. Blei}, {and} \bibinfo{person}{Joshua~B.
  Tenenbaum}.} \bibinfo{year}{2005}\natexlab{}.
\newblock \showarticletitle{{Integrating topics and syntax}}.
\newblock \bibinfo{journal}{\emph{Advances in Neural Information Processing
  Systems}} (\bibinfo{year}{2005}).
\newblock
\showISBNx{0262195348}
\showISSN{10495258}


\bibitem[Grimmer and Stewart(2013)]%
        {Grimmer2013}
\bibfield{author}{\bibinfo{person}{Justin Grimmer} {and}
  \bibinfo{person}{Brandon~M. Stewart}.} \bibinfo{year}{2013}\natexlab{}.
\newblock \showarticletitle{{Text as data: The promise and pitfalls of
  automatic content analysis methods for political texts}}.
\newblock \bibinfo{journal}{\emph{Political Analysis}} \bibinfo{volume}{21},
  \bibinfo{number}{3} (\bibinfo{year}{2013}), \bibinfo{pages}{267--297}.
\newblock
\showISSN{14764989}
\urldef\tempurl%
\url{https://doi.org/10.1093/pan/mps028}
\showDOI{\tempurl}


\bibitem[Holland et~al\mbox{.}(1983)]%
        {Holland1983}
\bibfield{author}{\bibinfo{person}{Paul~W. Holland},
  \bibinfo{person}{Kathryn~Blackmond Laskey}, {and} \bibinfo{person}{Samuel
  Leinhardt}.} \bibinfo{year}{1983}\natexlab{}.
\newblock \showarticletitle{{Stochastic blockmodels: First steps}}.
\newblock \bibinfo{journal}{\emph{Social Networks}} \bibinfo{volume}{5},
  \bibinfo{number}{2} (\bibinfo{year}{1983}), \bibinfo{pages}{109--137}.
\newblock
\showISSN{03788733}
\urldef\tempurl%
\url{https://doi.org/10.1016/0378-8733(83)90021-7}
\showDOI{\tempurl}


\bibitem[Hyland et~al\mbox{.}(2021)]%
        {Hyland2021}
\bibfield{author}{\bibinfo{person}{Charles~C Hyland}, \bibinfo{person}{Yuanming
  Tao}, \bibinfo{person}{Lamiae Azizi}, \bibinfo{person}{Martin Gerlach},
  \bibinfo{person}{Tiago~P Peixoto}, {and} \bibinfo{person}{Eduardo~G
  Altmann}.} \bibinfo{year}{2021}\natexlab{}.
\newblock \showarticletitle{{Multilayer networks for text analysis with
  multiple data types}}.
\newblock \bibinfo{journal}{\emph{EPJ Data Science}} \bibinfo{volume}{10},
  \bibinfo{number}{1} (\bibinfo{year}{2021}).
\newblock
\showISSN{21931127}
\urldef\tempurl%
\url{https://doi.org/10.1140/epjds/s13688-021-00288-5}
\showDOI{\tempurl}


\bibitem[Jiang and Brubaker(2018)]%
        {Jiang2018a}
\bibfield{author}{\bibinfo{person}{Jialun~"Aaron" Jiang} {and}
  \bibinfo{person}{Jed~R. Brubaker}.} \bibinfo{year}{2018}\natexlab{}.
\newblock \showarticletitle{{Tending Unmarked Graves}}.
\newblock \bibinfo{journal}{\emph{Proceedings of the ACM on Human-Computer
  Interaction}} \bibinfo{volume}{2}, \bibinfo{number}{CSCW}
  (\bibinfo{date}{nov} \bibinfo{year}{2018}), \bibinfo{pages}{1--19}.
\newblock
\showISSN{2573-0142}
\urldef\tempurl%
\url{https://doi.org/10.1145/3274350}
\showDOI{\tempurl}


\bibitem[Jiang et~al\mbox{.}(2021)]%
        {Jiang2021}
\bibfield{author}{\bibinfo{person}{Jialun~Aaron Jiang},
  \bibinfo{person}{Kandrea Wade}, \bibinfo{person}{Casey Fiesler}, {and}
  \bibinfo{person}{Jed~R. Brubaker}.} \bibinfo{year}{2021}\natexlab{}.
\newblock \showarticletitle{Supporting Serendipity: Opportunities and
  Challenges for Human-AI Collaboration in Qualitative Analysis}.
\newblock \bibinfo{journal}{\emph{Proc. ACM Hum.-Comput. Interact.}}
  \bibinfo{volume}{5}, \bibinfo{number}{CSCW1}, Article \bibinfo{articleno}{94}
  (\bibinfo{date}{apr} \bibinfo{year}{2021}), \bibinfo{numpages}{23}~pages.
\newblock
\urldef\tempurl%
\url{https://doi.org/10.1145/3449168}
\showDOI{\tempurl}


\bibitem[Kim et~al\mbox{.}(2020)]%
        {Kim2020}
\bibfield{author}{\bibinfo{person}{Hannah Kim}, \bibinfo{person}{Barry Drake},
  \bibinfo{person}{Alex Endert}, {and} \bibinfo{person}{Haesun Park}.}
  \bibinfo{year}{2020}\natexlab{}.
\newblock \showarticletitle{{ArchiText: Interactive Hierarchical Topic
  Modeling}}.
\newblock \bibinfo{journal}{\emph{IEEE Transactions on Visualization and
  Computer Graphics}} \bibinfo{volume}{2626}, \bibinfo{number}{c}
  (\bibinfo{year}{2020}).
\newblock
\showISSN{19410506}
\urldef\tempurl%
\url{https://doi.org/10.1109/TVCG.2020.2981456}
\showDOI{\tempurl}


\bibitem[Kivel{\"{a}} et~al\mbox{.}(2014)]%
        {Kivela2014}
\bibfield{author}{\bibinfo{person}{Mikko Kivel{\"{a}}}, \bibinfo{person}{Alex
  Arenas}, \bibinfo{person}{Marc Barthelemy}, \bibinfo{person}{James~P.
  Gleeson}, \bibinfo{person}{Yamir Moreno}, {and} \bibinfo{person}{Mason~A.
  Porter}.} \bibinfo{year}{2014}\natexlab{}.
\newblock \showarticletitle{{Multilayer networks}}.
\newblock \bibinfo{journal}{\emph{Journal of Complex Networks}}
  \bibinfo{volume}{2}, \bibinfo{number}{3} (\bibinfo{year}{2014}),
  \bibinfo{pages}{203--271}.
\newblock
\showISSN{20511329}
\urldef\tempurl%
\url{https://doi.org/10.1093/comnet/cnu016}
\showDOI{\tempurl}
\showeprint[arxiv]{1309.7233}


\bibitem[Klein et~al\mbox{.}(2015)]%
        {Klein2015}
\bibfield{author}{\bibinfo{person}{Lauren~F. Klein}, \bibinfo{person}{Jacob
  Eisenstein}, {and} \bibinfo{person}{Iris Sun}.}
  \bibinfo{year}{2015}\natexlab{}.
\newblock \showarticletitle{{Exploratory thematic analysis for digitized
  archival collections}}.
\newblock \bibinfo{journal}{\emph{Digital Scholarship in the Humanities}}
  \bibinfo{volume}{30}, \bibinfo{number}{October} (\bibinfo{year}{2015}),
  \bibinfo{pages}{i130--i141}.
\newblock
\showISSN{2055768X}
\urldef\tempurl%
\url{https://doi.org/10.1093/llc/fqv052}
\showDOI{\tempurl}


\bibitem[Lee et~al\mbox{.}(2012)]%
        {Lee2012a}
\bibfield{author}{\bibinfo{person}{Hanseung Lee}, \bibinfo{person}{Jaeyeon
  Kihm}, \bibinfo{person}{Jaegul Choo}, \bibinfo{person}{John Stasko}, {and}
  \bibinfo{person}{Haesun Park}.} \bibinfo{year}{2012}\natexlab{}.
\newblock \showarticletitle{{iVisClustering: An Interactive Visual Document
  Clustering via Topic Modeling}}.
\newblock \bibinfo{journal}{\emph{Computer Graphics Forum}}
  \bibinfo{volume}{31}, \bibinfo{number}{3pt3} (\bibinfo{year}{2012}),
  \bibinfo{pages}{1155--1164}.
\newblock
\showISSN{0167-7055}
\urldef\tempurl%
\url{https://doi.org/10.1111/j.1467-8659.2012.03108.x}
\showDOI{\tempurl}


\bibitem[Lee et~al\mbox{.}(2017)]%
        {Lee2017}
\bibfield{author}{\bibinfo{person}{Tak~Yeon Lee}, \bibinfo{person}{Alison
  Smith}, \bibinfo{person}{Kevin Seppi}, \bibinfo{person}{Niklas Elmqvist},
  \bibinfo{person}{Jordan Boyd-Graber}, {and} \bibinfo{person}{Leah
  Findlater}.} \bibinfo{year}{2017}\natexlab{}.
\newblock \showarticletitle{{The human touch: How non-expert users perceive,
  interpret, and fix topic models}}.
\newblock \bibinfo{journal}{\emph{International Journal of Human-Computer
  Studies}}  \bibinfo{volume}{105} (\bibinfo{year}{2017}),
  \bibinfo{pages}{28--42}.
\newblock
\showISSN{1071-5819}
\urldef\tempurl%
\url{https://doi.org/10.1016/j.ijhcs.2017.03.007}
\showDOI{\tempurl}


\bibitem[McDonald et~al\mbox{.}(2019)]%
        {McDonald2019}
\bibfield{author}{\bibinfo{person}{Nora McDonald}, \bibinfo{person}{Sarita
  Schoenebeck}, {and} \bibinfo{person}{Andrea Forte}.}
  \bibinfo{year}{2019}\natexlab{}.
\newblock \showarticletitle{{Reliability and Inter-rater Reliability in
  Qualitative Research}}.
\newblock \bibinfo{journal}{\emph{Proceedings of the ACM on Human-Computer
  Interaction}} \bibinfo{volume}{3}, \bibinfo{number}{CSCW}
  (\bibinfo{year}{2019}), \bibinfo{pages}{1--23}.
\newblock
\urldef\tempurl%
\url{https://doi.org/10.1145/3359174}
\showDOI{\tempurl}


\bibitem[Miles et~al\mbox{.}(2013)]%
        {miles2013qualitative}
\bibfield{author}{\bibinfo{person}{M.B. Miles}, \bibinfo{person}{A.M.
  Huberman}, {and} \bibinfo{person}{J. Saldana}.}
  \bibinfo{year}{2013}\natexlab{}.
\newblock \bibinfo{booktitle}{\emph{Qualitative Data Analysis: A Methods
  Sourcebook}}.
\newblock \bibinfo{publisher}{SAGE Publications}.
\newblock
\showISBNx{9781483323794}
\urldef\tempurl%
\url{https://books.google.com/books?id=p0wXBAAAQBAJ}
\showURL{%
\tempurl}


\bibitem[Mimno and Blei(2011)]%
        {Mimno2011}
\bibfield{author}{\bibinfo{person}{David Mimno} {and} \bibinfo{person}{David
  Blei}.} \bibinfo{year}{2011}\natexlab{}.
\newblock \showarticletitle{{Bayesian checking for topic models}}.
\newblock \bibinfo{journal}{\emph{EMNLP 2011 - Conference on Empirical Methods
  in Natural Language Processing, Proceedings of the Conference}}
  (\bibinfo{year}{2011}), \bibinfo{pages}{227--237}.
\newblock
\showISBNx{1937284115}


\bibitem[Muller et~al\mbox{.}(2016)]%
        {Muller2016}
\bibfield{author}{\bibinfo{person}{Michael Muller}, \bibinfo{person}{Shion
  Guha}, \bibinfo{person}{Eric~P.S. Baumer}, \bibinfo{person}{David Mimno},
  {and} \bibinfo{person}{N~Sadat Shami}.} \bibinfo{year}{2016}\natexlab{}.
\newblock \showarticletitle{{Machine Learning and Grounded Theory Method}}. In
  \bibinfo{booktitle}{\emph{Proceedings of the 19th International Conference on
  Supporting Group Work}}. \bibinfo{publisher}{ACM}, \bibinfo{address}{New
  York, NY, USA}, \bibinfo{pages}{3--8}.
\newblock
\showISBNx{9781450342766}
\urldef\tempurl%
\url{https://doi.org/10.1145/2957276.2957280}
\showDOI{\tempurl}


\bibitem[Nelson(2020)]%
        {Nelson2020}
\bibfield{author}{\bibinfo{person}{Laura~K. Nelson}.}
  \bibinfo{year}{2020}\natexlab{}.
\newblock \showarticletitle{{Computational Grounded Theory: A Methodological
  Framework}}.
\newblock \bibinfo{journal}{\emph{Sociological Methods and Research}}
  \bibinfo{volume}{49}, \bibinfo{number}{1} (\bibinfo{year}{2020}),
  \bibinfo{pages}{3--42}.
\newblock
\showISSN{15528294}
\urldef\tempurl%
\url{https://doi.org/10.1177/0049124117729703}
\showDOI{\tempurl}


\bibitem[Paul and Girju(2010)]%
        {Paul2010}
\bibfield{author}{\bibinfo{person}{Michael Paul} {and} \bibinfo{person}{Roxana
  Girju}.} \bibinfo{year}{2010}\natexlab{}.
\newblock \showarticletitle{{A two-dimensional Topic-Aspect Model for
  discovering multi-faceted topics}}.
\newblock \bibinfo{journal}{\emph{Proceedings of the National Conference on
  Artificial Intelligence}}  \bibinfo{volume}{1} (\bibinfo{year}{2010}),
  \bibinfo{pages}{545--550}.
\newblock
\showISBNx{9781577354642}


\bibitem[Peixoto(2019)]%
        {Peixoto2019}
\bibfield{author}{\bibinfo{person}{Tiago~P. Peixoto}.}
  \bibinfo{year}{2019}\natexlab{}.
\newblock \showarticletitle{{Bayesian stochastic blockmodeling}}.
\newblock \bibinfo{journal}{\emph{Advances in Network Clustering and
  Blockmodeling}} (\bibinfo{year}{2019}), \bibinfo{pages}{289--332}.
\newblock
\showISBNx{9781119483298}
\urldef\tempurl%
\url{https://doi.org/10.1002/9781119483298.ch11}
\showDOI{\tempurl}
\showeprint[arxiv]{1705.10225}


\bibitem[Steyvers and Griffiths(2010)]%
        {Steyvers2010}
\bibfield{author}{\bibinfo{person}{M Steyvers} {and} \bibinfo{person}{T
  Griffiths}.} \bibinfo{year}{2010}\natexlab{}.
\newblock \showarticletitle{{Probalistic Topic Models}}.
\newblock \bibinfo{journal}{\emph{Latent Semantic Analysis: A Road To Meaning}}
  \bibinfo{volume}{3}, \bibinfo{number}{3} (\bibinfo{year}{2010}),
  \bibinfo{pages}{993--1022}.
\newblock
\showISBNx{1532-4435}
\showISSN{1386-4564}
\showeprint[arxiv]{1111.6189v1}
\urldef\tempurl%
\url{http://www.sciencedirect.com/science/article/pii/S0140366413001047{\%}5Cnhttp://ceas.cc/2004/167.pdf{\%}5Cnhttp://doi.acm.org/10.1145/1806338.1806450{\%}5Cnhttp://eprints.soton.ac.uk/272254/{\%}5Cnhttp://ieeexplore.ieee.org/lpdocs/epic03/wrapper.htm?arnumber=7033160{\%}25}
\showURL{%
\tempurl}


\bibitem[Tversky(1991)]%
        {tversky1991spatial}
\bibfield{author}{\bibinfo{person}{Barbara Tversky}.}
  \bibinfo{year}{1991}\natexlab{}.
\newblock \showarticletitle{Spatial mental models}.
\newblock \bibinfo{journal}{\emph{Psychology of Learning and Motivation}}
  \bibinfo{volume}{27} (\bibinfo{year}{1991}), \bibinfo{pages}{109--145}.
\newblock


\bibitem[Wiedemann(2013)]%
        {wiedemann2013opening}
\bibfield{author}{\bibinfo{person}{Gregor Wiedemann}.}
  \bibinfo{year}{2013}\natexlab{}.
\newblock \showarticletitle{Opening up to big data: Computer-assisted analysis
  of textual data in social sciences}.
\newblock \bibinfo{journal}{\emph{Historical Social Research/Historische
  Sozialforschung}} (\bibinfo{year}{2013}), \bibinfo{pages}{332--357}.
\newblock


\bibitem[Xie and Xing(2013)]%
        {Xie2013}
\bibfield{author}{\bibinfo{person}{Pengtao Xie} {and} \bibinfo{person}{Eric~P.
  Xing}.} \bibinfo{year}{2013}\natexlab{}.
\newblock \showarticletitle{{Integrating document clustering and topic
  modeling}}.
\newblock \bibinfo{journal}{\emph{Uncertainty in Artificial Intelligence -
  Proceedings of the 29th Conference, UAI 2013}} (\bibinfo{year}{2013}),
  \bibinfo{pages}{694--703}.
\newblock
\showeprint[arxiv]{1309.6874}


\bibitem[Yang et~al\mbox{.}(2015)]%
        {Yang2015a}
\bibfield{author}{\bibinfo{person}{Yi Yang}, \bibinfo{person}{Doug Downey},
  {and} \bibinfo{person}{Jordan Boyd-Graber}.} \bibinfo{year}{2015}\natexlab{}.
\newblock \showarticletitle{{Efficient methods for incorporating knowledge into
  topic models}}. In \bibinfo{booktitle}{\emph{Conference Proceedings - EMNLP
  2015: Conference on Empirical Methods in Natural Language Processing}}.
  \bibinfo{publisher}{Association for Computational Linguistics},
  \bibinfo{pages}{308--317}.
\newblock
\showISBNx{9781941643327}
\urldef\tempurl%
\url{https://doi.org/10.18653/v1/d15-1037}
\showDOI{\tempurl}


\end{thebibliography}

\appendix

\section{Recipes Dataset}
\label{sec:recipes}
All recipe posts on Simply Recipes (https://www.simplyrecipes.com) were crawled and retrieved on November 8th, 2021. 
The corpus consists of 2,615 recipes with 14,343 unique words after lemmatization and removal of all words (using the spacy package) except content words (proper nouns, adjectives, adverbs, nouns, verbs).
This resulted in 720,292 total edges in the Text layer, with the average document length (node-degree) of 275.45.

\end{document}